\documentclass[preprint,showpacs,preprintnumbers,superscriptaddress,aps]{revtex4}
\usepackage[colorlinks,linkcolor=blue,anchorcolor=blue,citecolor=blue]{hyperref}
\usepackage{amsmath}
\usepackage{amssymb}
\usepackage{amsfonts}
\usepackage{subfigure}
\usepackage{graphicx}
\usepackage{dcolumn}
\usepackage{bm}

\begin{document}
	
	\title{Hydrodynamic Assessment of Direct Drive Inertial Confinement Fusion with Mixed $2\omega-3\omega$ Lasers}
	
	\author{Guannan Zheng}
	\affiliation{Department of Plasma Physics and Fusion Engineering, University of Science and Technology of China, Hefei, Anhui 230026, People's Republic of China}
	
	\author{Tao Tao}
	\affiliation{Department of Plasma Physics and Fusion Engineering, University of Science and Technology of China, Hefei, Anhui 230026, People's Republic of China}
	
	\author{Qing Jia}
	\affiliation{Department of Plasma Physics and Fusion Engineering, University of Science and Technology of China, Hefei, Anhui 230026, People's Republic of China}
	
	\author{Jun Li}
    \affiliation{Department of Plasma Physics and Fusion Engineering, University of Science and Technology of China, Hefei, Anhui 230026, People's Republic of China}
	
	\author{Rui Yan}
	\affiliation{Department of Modern Mechanics, University of Science and Technology of China, Hefei, Anhui 230026, People's Republic of China}
	\affiliation{Collaborative Innovation Center of IFSA, Shanghai Jiao Tong University, Shanghai 200240, People's Republic of China}
	
    \author{Jian Zheng}
	\email[To whom correspondence should be addressed: ]{jzheng@ustc.edu.cn}
	\affiliation{Department of Plasma Physics and Fusion Engineering, University of Science and Technology of China, Hefei, Anhui 230026, People's Republic of China}
	\affiliation{Collaborative Innovation Center of IFSA, Shanghai Jiao Tong University, Shanghai 200240, People's Republic of China}
	
\begin{abstract}
Ablation with mixed $2\omega$--$3\omega$ lasers is investigated as a possible drive strategy for balancing drive efficiency and ablative stabilization in direct-drive inertial confinement fusion. One-dimensional radiation-hydrodynamic simulations are performed for planar CH targets using the FLASH code [B. Fryxell et al, The Astrophysical Journal Supplement Series \textbf{131}, 273 (2000)]. The total target-incident laser intensity is varied from 100 to $1600~\mathrm{TW}/\mathrm{cm}^{2}$, and the $3\omega$ laser intensity fraction is scanned from 0 to 100\%. Thick-target simulations are used to determine quasi-steady ablation-pressure scalings, while thin-foil simulations are used to characterize the acceleration stage and to evaluate the linear ablative Rayleigh--Taylor instability (RTI) gain using a Takabe-type model. The simulations show that adding a $3\omega$ component to a $2\omega$-dominated drive increases the effective ablation pressure, enhances the ablation velocity, and reduces the maximum linear RTI gain. Within the present one-dimensional hydrodynamic model, the mixed drive also reduces the target-incident energy required to accelerate the foil to $300~\mathrm{km}/\mathrm{s}$, especially at high intensity. This improvement is attributed to the deeper penetration of $3\omega$ light, which deposits energy closer to the dense ablation region and enhances conductive heat transport toward the ablation front. These results suggest that mixed-wavelength drive can recover much of the favorable hydrodynamic performance of $3\omega$ irradiation while retaining part of the energy-accessibility advantage of $2\omega$ operation, providing an additional design space of freedom for direct-drive target optimization.
\end{abstract}

\maketitle

\section{Introduction}
\label{Introduction}

Inertial confinement fusion (ICF) aims to initiate thermonuclear burn by compressing and heating a small fuel capsule with intense laser or particle drivers \cite{Nuckolls1972,Craxton2015,Betti2016}. Recent ignition and target-gain experiments at the National Ignition Facility (NIF) have demonstrated ignition-scale fusion in the laboratory and represent a major milestone in high-energy-density physics \cite{AbuShawareb2022,Kritcher2022,Zylstra2022,AbuShawareb2024}. For inertial fusion energy, however, ignition must be accompanied by high target gain, high driver efficiency, hydrodynamic robustness, and target designs compatible with high-repetition-rate operation. These requirements have motivated renewed interest in direct-drive ICF, in which laser beams irradiate the capsule directly rather than first converting their energy into X rays inside a hohlraum \cite{Craxton2015,Hurricane2023}. By avoiding the intermediate laser-to-X-ray conversion, direct drive offers a simpler target configuration and a potentially higher laser-to-capsule coupling efficiency than indirect drive. This advantage comes with more stringent requirements on irradiation uniformity, laser imprint, shock control, and acceleration-stage hydrodynamic stability, especially the ablative Rayleigh--Taylor instability (ARTI).

The laser wavelength is a key control parameter in direct-drive target design. Present ignition-scale Nd:glass laser facilities commonly use the third harmonic, $3\omega$, with a wavelength of approximately $0.35~\mu\mathrm{m}$~\cite{10.1063/1.1578638,BOEHLY1997495,FLEUROT2005147,Miquel_2016,LIN199961,SGIII_2016}. Shorter-wavelength irradiation has several well-known hydrodynamic advantages for ICF. The critical density scales as
\begin{equation}
    n_c \propto \omega^2 ,
\end{equation}
so that $3\omega$ light can propagate into denser plasma than $2\omega$ light. As a result, the laser energy is deposited closer to the ablation front, which generally increases the ablation pressure and the mass ablation rate under comparable drive conditions \cite{Schmitt2023,Lu2025}. These properties are favorable for shell acceleration and for the mitigation of ablative RTI growth. In a commonly used Takabe-type form, the ablative RTI growth rate is written as \cite{Takabe1985,Betti1998}
\begin{equation}
    \gamma =
    \alpha
    \sqrt{\frac{k g}{1+kL_m}}
    - \beta k V_a ,
    \label{eq:takabe_growth}
\end{equation}
where $k$ is the perturbation wavenumber, $g$ is the shell acceleration, $L_m$ is the density-gradient scale length near the ablation front, and $V_a$ is the ablation velocity. A larger $V_a$ strengthens the ablative stabilization term, whereas a larger $L_m$ reduces the effective RTI drive, especially for short-wavelength perturbations. From the viewpoint of acceleration-stage hydrodynamic stability, $3\omega$ drive is therefore generally advantageous.

Exclusive reliance on $3\omega$ irradiation, however, introduces system-level constraints. Frequency conversion to the ultraviolet reduces the laser energy available at the target, and ultraviolet optical components typically have lower damage thresholds than longer-wavelength optics. Consequently, the maximum deliverable fluence, energy, or power of a large laser facility can be more restrictive in $3\omega$ operation than in longer-wavelength operation \cite{Suter2004NF,Suter2004POP,Wilson2021}. Operation at the second harmonic, $2\omega$, with a wavelength of approximately $0.53~\mu\mathrm{m}$, can in principle relax some of these constraints in certain laser architectures, owing in part to the higher damage tolerance of longer-wavelength optics. The larger accessible laser energy may enable designs with larger capsules, thicker ablators, lower in-flight aspect ratios, or improved hydrodynamic robustness \cite{Strobel2004,Atzeni2012}.

The hydrodynamic drawback of $2\omega$ drive is that its lower critical density shifts the laser absorption region farther away from the ablation front. The longer thermal-conduction path can reduce the efficiency with which absorbed laser energy is converted into ablation pressure. In addition, longer-wavelength drive typically produces a lower ablation velocity, which weakens ablative stabilization during the acceleration phase \cite{Schmitt2023,Lu2025}. Longer wavelengths can also increase the susceptibility to laser--plasma instabilities (LPIs), including stimulated Raman scattering, stimulated Brillouin scattering, two-plasmon decay \cite{Stevenson2004,Stevenson2005,Niemann2005,Moody2009,Depierreux2009,Depierreux2012,Wasser2024}, etc. Although such effects may be mitigated by beam smoothing, plasma conditioning, laser bandwidth, and favorable high-temperature damping \cite{Niemann2008,Li2012,Hume2025}, they remain important considerations for integrated target designs. The present work focuses on the hydrodynamic consequences of wavelength choice and should therefore be regarded as a hydrodynamic assessment rather than a complete evaluation including multidimensional LPI and imprint physics.

These considerations reveal a trade-off in single-wavelength direct drive. The $3\omega$ drive provides favorable ablation physics, higher ablation pressure, and stronger ablative RTI stabilization, but it is more constrained by frequency conversion and ultraviolet-optics limitations. The $2\omega$ drive may provide access to a larger laser-energy reservoir and, in some cases, a longer smoothing distance for laser-imprint perturbations, but it generally gives lower target hydrodynamic coupling and weaker ablative stabilization. A single-wavelength strategy therefore has difficulty optimizing energy availability, ablation efficiency, imprint smoothing, and RTI mitigation simultaneously.

A mixed-wavelength drive provides a possible route to relax this trade-off. In a mixed $2\omega$--$3\omega$ direct-drive scheme, the two laser components deposit energy at different locations in the coronal plasma because of their different critical densities. The $2\omega$ component is absorbed farther out in the lower-density corona, whereas the $3\omega$ component penetrates deeper and deposits energy closer to the ablation front. This two-region deposition structure can modify the coronal temperature profile, electron heat flux, ablation pressure, ablation velocity, and density-gradient scale length in a way that cannot, in general, be represented by a single effective wavelength. In particular, a partial $3\omega$ component may improve the coupling of thermal energy to the dense ablation region, thereby recovering much of the hydrodynamic performance and RTI-stabilizing effect of pure $3\omega$ irradiation while retaining part of the energy-accessibility advantage associated with $2\omega$ operation.

The key questions are how the mixed drive modifies the ablation-pressure scaling, whether the resulting changes in $V_a$ and $L_m$ reduce the linear RTI gain, and how the target-incident laser energy required for a prescribed foil velocity depends on the mixing ratio. In this paper, we address these questions using one-dimensional (1D) planar radiation-hydrodynamic simulations with the FLASH code \cite{Fryxell_2000}. The total target-incident laser intensity is varied over a broad range, and the fraction of the intensity carried by the $3\omega$ component is scanned from pure $2\omega$ to pure $3\omega$. Two CH target configurations are considered. A $600~\mu\mathrm{m}$-thick CH target is used to characterize the quasi-steady ablation regime and to obtain ablation-pressure scalings. A $100~\mu\mathrm{m}$-thick CH foil is used to study the acceleration stage and to extract the hydrodynamic quantities relevant to ablative RTI, including the shell acceleration $g$, the ablation velocity $V_a$, and the density-gradient scale length $L_m$.

The objectives of this work are threefold. First, we quantify the ablation-pressure scaling under mixed $2\omega$--$3\omega$ irradiation and compare the thick-target quasi-steady regime with the finite-foil acceleration regime. Second, we use a Takabe-type linear stability model to evaluate how the mixed drive changes the maximum RTI gain through its effects on $g$, $V_a$, and $L_m$. Third, we estimate the target-incident laser energy required to accelerate the foil to $300~\mathrm{km/s}$ and analyze the underlying energy-transport mechanism in terms of critical-surface location, electron temperature, thermal conductivity, and conductive heat flux. Particular attention is paid to intermediate mixing ratios, which may provide a practical hydrodynamic compromise between the benefits of short-wavelength drive and the energy accessibility of longer-wavelength operation. The remainder of this paper is organized as follows. Section~\ref{sec:model} describes the simulation model, target configurations, laser-energy deposition model, and definitions of the extracted hydrodynamic quantities. Section~\ref{sec:pressure} presents the ablation-pressure scaling in the quasi-steady and acceleration regimes. Section~\ref{sec:rti_model} introduces the RTI gain-envelope analysis, and Sec.~\ref{sec:rti_results} applies it to the simulation results. Section~\ref{sec:energy} discusses the target-incident laser-energy requirement and the associated thermal-transport mechanism. Section~\ref{sec:discussion} discusses the design implications of mixed-wavelength drive, and Sec.~\ref{sec:conclusion} summarizes the conclusions.

\section{Simulation model}
\label{sec:model}

The hydrodynamic response of CH targets driven by mixed $2\omega$-$3\omega$ lasers is investigated using one-dimensional planar radiation-hydrodynamic simulations with the FLASH code \cite{Fryxell_2000}. In the code, radiation transport is treated with a multigroup radiation-diffusion model using 20 photon-energy groups. The opacity data are taken from the SNOP database, and the CH equation of state is generated from the QEOS model \cite{10.1063/1.866963,KEMP1998674}. The initial CH target has a uniform density of
\begin{equation}
\rho_0 = 1.06~\mathrm{g/cm^3},
\end{equation}
and an initial temperature of
\begin{equation}
T_0 = 300~\mathrm{K}.
\end{equation}
Two target configurations are considered. A $600~\mu\mathrm{m}$-thick CH target is used to study the steady ablation regime and to determine the ablation-pressure scaling. A $100~\mu\mathrm{m}$-thick CH foil is used to analyze the acceleration stage and to extract the hydrodynamic parameters relevant to RTI. Outflow conditions are applied at both the left and right boundaries. Laser is injected from the right boundary and propagates in the negative $x$-direction toward the target. The computational domain $\left[0, 2000~\mu\mathrm{m}\right]$, with 4096 grid cells ($\Delta x \sim 0.5~\mu\mathrm{m}$), is sufficiently large to include the expanding coronal plasma, the laser absorption region, the ablation front, and the accelerated shell.

Laser energy deposition is calculated through inverse bremsstrahlung absorption using the ray-tracing package implemented in FLASH. Although the ray-tracing model can describe laser refraction in multidimensional geometry, the present study is restricted to 1D planar simulations. Laser rays are not propagated beyond their corresponding critical surfaces. Since the critical density satisfies
\begin{equation}
n_c = \frac{m_e \epsilon_0 \omega^2}{e^2},
\end{equation}
where $\omega$ is the laser angular frequency, one has
\begin{equation}
n_c \propto \omega^2 .
\end{equation}
Therefore, the critical density of the $3\omega$ laser is higher than that of the $2\omega$ laser:
\begin{equation}
\frac{n_{c,3\omega}}{n_{c,2\omega}}
=
\left( \frac{3\omega}{2\omega} \right)^2
=
\frac{9}{4}.
\end{equation}
As a result, the $3\omega$ component can penetrate into a denser plasma region and deposit energy closer to the ablation front, while the $2\omega$ component is absorbed farther out in the lower-density corona.

Electron thermal conduction is modeled using the classical Spitzer--H\"arm heat flux with a Larsen flux limiter. The heat flux is written as
\begin{equation}
q = -D_{\mathrm{fl}}\nabla T_e ,
\end{equation}
where the flux-limited diffusion coefficient is
\begin{equation}
D_{\mathrm{fl}}
=
\frac{1}
{
\left[
\left( \frac{1}{D_{\mathrm{SH}}} \right)^2
+
\left(
\frac{|\nabla T_e|}{q_{\max}}
\right)^2
\right]^{1/2}
}.
\end{equation}
Here $D_{\mathrm{SH}}$ is the classical Spitzer--H\"arm diffusion coefficient. The maximum heat flux is given by
\begin{equation}
q_{\max} = f_{\mathrm{lim}} q_{\mathrm{fs}},
\end{equation}
where $q_{\mathrm{fs}}$ is the free-streaming heat flux. In this work, the flux-limiter coefficient is fixed at
\begin{equation}
f_{\mathrm{lim}} = 0.06 .
\end{equation}
This treatment provides a smooth transition between the classical conduction regime and the flux-limited regime in steep temperature-gradient regions.

The mixed-wavelength drive is specified by the total laser intensity
\begin{equation}
I_0 = I_{2\omega} + I_{3\omega},
\end{equation}
and the $3\omega$ fraction
\begin{equation}
f_{3\omega}
=
\frac{I_{3\omega}}
{I_{2\omega}+I_{3\omega}} .
\end{equation}
Thus,
\begin{equation}
I_{3\omega} = f_{3\omega} I_0,
\end{equation}
and
\begin{equation}
I_{2\omega} = (1-f_{3\omega}) I_0 .
\end{equation}
The cases $f_{3\omega}=0$ and $f_{3\omega}=1$ correspond to pure $2\omega$ and pure $3\omega$ drive, respectively. An equal-intensity mixed drive, defined by
\begin{equation}
I_{2\omega}:I_{3\omega} = 1:1,
\end{equation}
corresponds to $f_{3\omega}=0.5$. In the comparison of different mixing ratios, the total incident laser intensity is kept fixed, so that the effect of redistributing laser energy between the two wavelengths can be isolated.

\begin{figure}[htbp]
    \centering
    \includegraphics[width=0.60\textwidth]{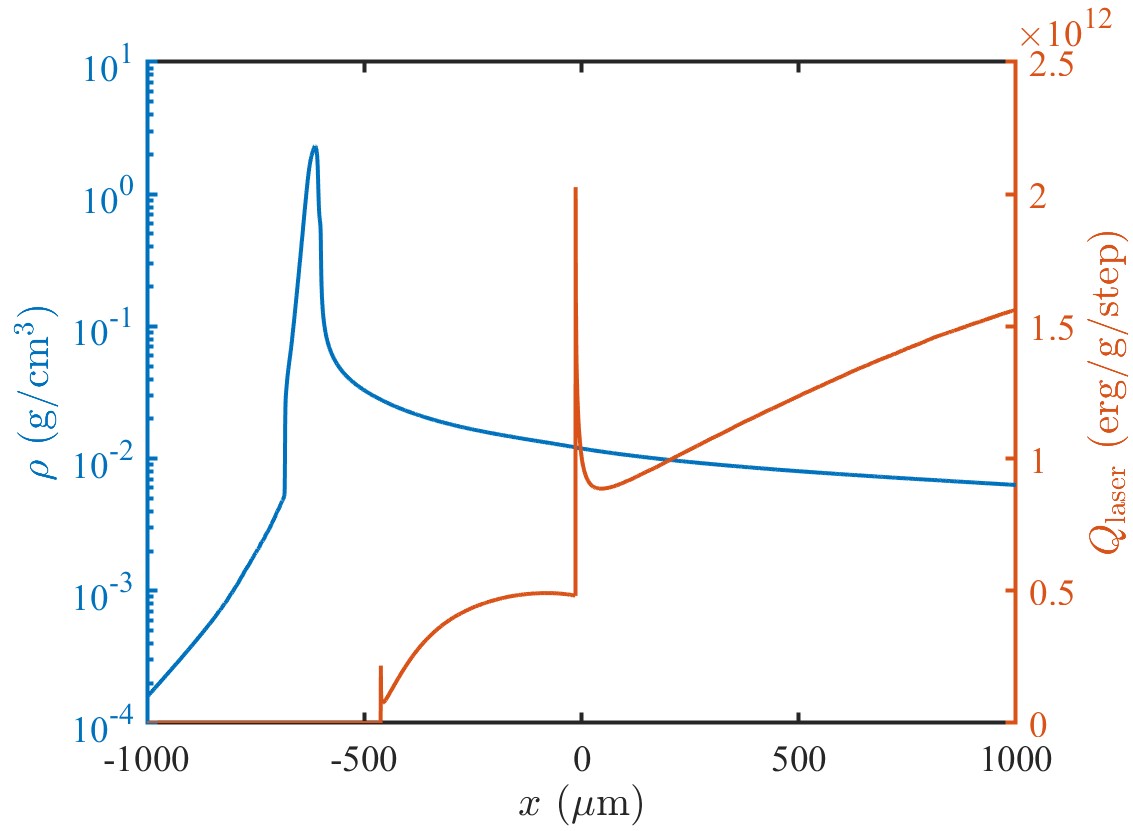}
    \caption{
    Density profile and laser-energy-deposition profile of a $100~\mu\mathrm{m}$ CH foil driven by an equal-intensity mixed $2\omega$-$3\omega$ laser, $I_{2\omega}:I_{3\omega}=1:1$. The total laser intensity is $I_0=1000~\mathrm{TW/cm^2}$, and the profiles are shown at $t=4~\mathrm{ns}$, when the target has reached the quasi-steady acceleration stage. The blue curve denotes the mass density, and the orange curve denotes the laser-energy-deposition rate. The spatial separation of the energy-deposition regions results from the different critical densities of the $2\omega$ and $3\omega$ laser components.
    }
    \label{fig:mixed_drive_profile}
\end{figure}

Figure~\ref{fig:mixed_drive_profile} shows a representative quasi-steady flow structure during the acceleration stage for the equal-intensity mixed $2\omega$-$3\omega$ drive. The total laser intensity is $I_0=1000~\mathrm{TW/cm^2}$, and the profiles are taken at $t=4~\mathrm{ns}$. The blue curve represents the mass density, and the orange curve represents the laser energy deposition rate. The density profile shows a low-density coronal plasma on the laser-irradiated side, a steep density rise near the ablation front, and a compressed dense shell behind the ablation front. Laser energy deposition exhibits two spatially separated regions because the $2\omega$ and $3\omega$ components have different critical densities. The outer deposition region is mainly associated with the $2\omega$ component, while the inner deposition region closer to the ablation front is mainly associated with the $3\omega$ component. This two-region deposition structure modifies the coronal temperature profile, the heat flux transported to the ablation front, and consequently the ablation pressure and the RTI-relevant hydrodynamic parameters.

For the steady-ablation calculations with the $600~\mu\mathrm{m}$ CH target, the hydrodynamic quantities are extracted at $t=5~\mathrm{ns}$, which is used as the reference time for the quasi-steady ablation flow. For the acceleration calculations with the $100~\mu\mathrm{m}$ CH foil, the reference state is chosen when the foil has moved by approximately
\begin{equation}
\Delta x = 600~\mu\mathrm{m}.
\end{equation}
This displacement-based criterion provides a consistent comparison between cases with different acceleration histories. Hydrodynamic quantities extracted from the simulations include the ablation pressure $P_a$, the ablation velocity $V_a$, the density-gradient scale length $L_m$, and the shell acceleration $g$. The ablation pressure is evaluated near the ablation front and is defined as the maximum pressure. Ablation velocity is defined as
\begin{equation}
V_a = \frac{\dot{m}}{\rho_a},
\end{equation}
where $\dot{m}$ is the mass ablation rate per unit area and $\rho_a$ is the density at the ablation front. The density-gradient scale length is defined by
\begin{equation}
L_m
=
\left|
\frac{\rho}
{\partial \rho/\partial x}
\right|_{a,\,\min}.
\end{equation}
The shell acceleration is calculated from
\begin{equation}
g = \frac{d v_{\mathrm{sh}}}{dt},
\end{equation}
where $v_{\mathrm{sh}}$ is the shell velocity, obtained consistently for all drive conditions from the motion of the accelerated dense shell.

\begin{figure}[htbp]
    \centering
    \includegraphics[width=\textwidth]{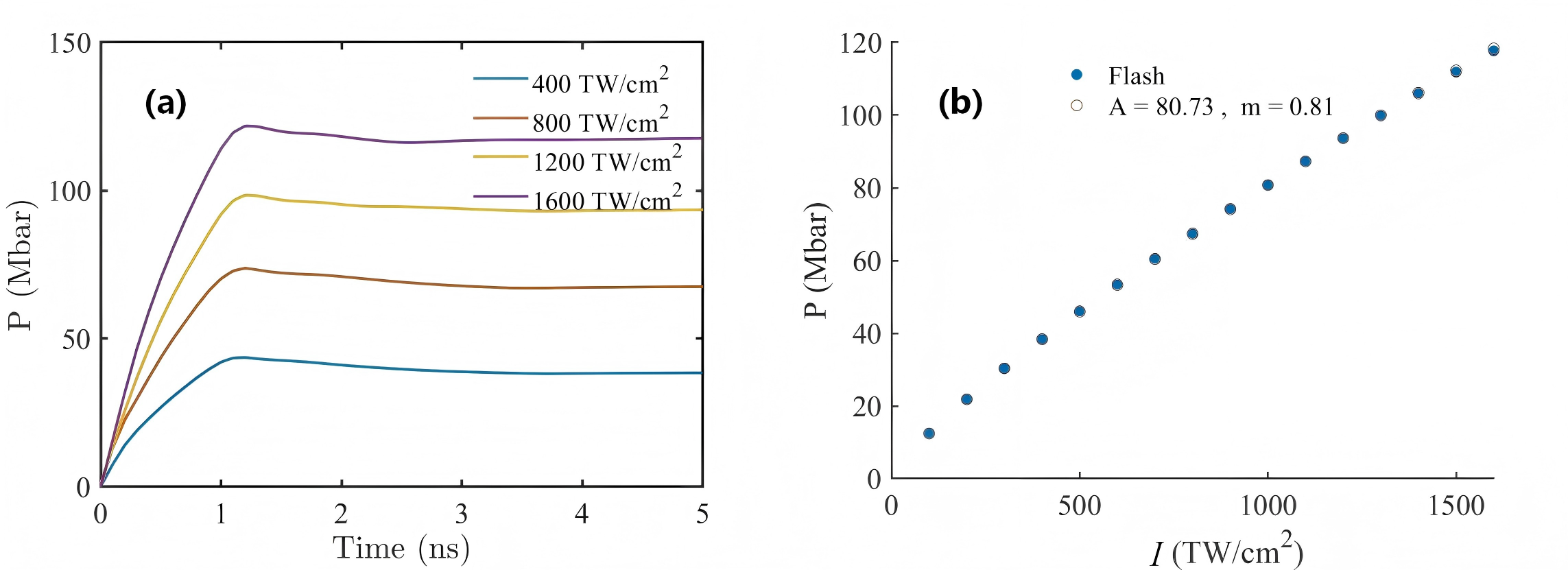}
    \caption{
    Quasi-steady ablation-pressure scaling for a $600~\mu\mathrm{m}$ CH target at $f_{3\omega}=50\%$. (a) Temporal evolution of the ablation pressure for laser intensities of $400$, $800$, $1200$, and $1600~\mathrm{TW/cm^2}$. The pressure rapidly rises during the initial transient and approaches a quasi-steady value by $t=5~\mathrm{ns}$. (b) Ablation pressure extracted at $t=5~\mathrm{ns}$ as a function of laser intensity. The FLASH simulation results are well fitted by $P_a=A I^m$, with $A=80.73$ and $m=0.81$, where $P_a$ is in Mbar and $I$ is in $\mathrm{PW/cm^2}$.
    }
    \label{fig:steady_scaling}
\end{figure}

\section{Ablation-pressure scaling}
\label{sec:pressure}

To quantify the effect of the mixed $2\omega$--$3\omega$ laser drive on the hydrodynamic response, we first examine the ablation-pressure scaling in the quasi-steady ablation regime and then extend the analysis to the foil-acceleration stage. In all simulations, the laser intensity is linearly ramped from zero to its peak value during the first 1 ns, after which it remains constant. The laser intensity is denoted by $I$, and the fraction of the total laser intensity carried by the $3\omega$ component is defined as $f_{3\omega}$. Unless otherwise specified, $I$ is expressed in $\mathrm{PW/cm^2}$ in the fitting formulas. The ablation pressure is fitted by a power-law form,
\begin{equation}
P_a = A I^m ,
\label{eq:Pa_scaling}
\end{equation}
where $P_a$ is in Mbar, $A$ is the normalization coefficient, and $m$ is the intensity-scaling exponent. In this paper, the dependence of $A$ and $m$ on $f_{3\omega}$ is obtained directly from the FLASH simulation data.

\subsection{Quasi-steady ablation in thick CH targets}

Figure~\ref{fig:steady_scaling}(a) shows the temporal evolution of the ablation pressure for a $600~\mu\mathrm{m}$ CH target at $f_{3\omega}=50\%$. Four representative laser intensities are considered: $400$, $800$, $1200$, and $1600~\mathrm{TW/cm^2}$. For all intensities, the pressure increases rapidly during the initial transient stage and then approaches a nearly time-independent value. The pressure reaches its maximum at approximately $1$--$1.5~\mathrm{ns}$, followed by a slight relaxation toward a quasi-steady plateau. By $t=5~\mathrm{ns}$, the pressure variation is small, indicating that the system has entered a quasi-steady ablation regime. Therefore, the ablation quantities at $t=5~\mathrm{ns}$ are used to characterize the steady-ablation state of the thick CH target. This choice avoids the initial transient response while providing a representative pressure value before any significant finite-target effects become relevant for the thick target. For $f_{3\omega}=50\%$, the pressure extracted at $t=5~\mathrm{ns}$ follows a clear power-law dependence on the laser intensity, as shown in Fig.~\ref{fig:steady_scaling}(b). The FLASH simulation data are accurately reproduced by
\begin{equation}
P_a(\mathrm{Mbar}) = 80.73 I^{0.81},
\label{eq:steady_scaling_f50}
\end{equation}
where $I$ is in $\mathrm{PW/cm^2}$. The fitted curve nearly overlaps with the simulation data over the entire intensity range considered here. This agreement indicates that, for a fixed mixing fraction, the ablation pressure under mixed-wavelength irradiation can still be described by a simple intensity power law. The fitted intensity exponent in the present steady-ablation simulations, $m\simeq0.80$--$0.82$, is close to the value $0.79$ reported by Schmitt and Obenschain~\cite{Schmitt2023} for the scaling $P_{\mathrm{a}}\propto I_{\mathrm{abs}}^{0.79}\lambda^{-0.28}$. Here, however, the present scaling is constructed with respect to the total incident intensity for mixed $2\omega/3\omega$ irradiation, whereas Ref.~\cite{Schmitt2023} used the absorbed intensity for single-wavelength irradiation.

\begin{figure}[htbp]
	\centering
	\includegraphics[width=0.60\textwidth]{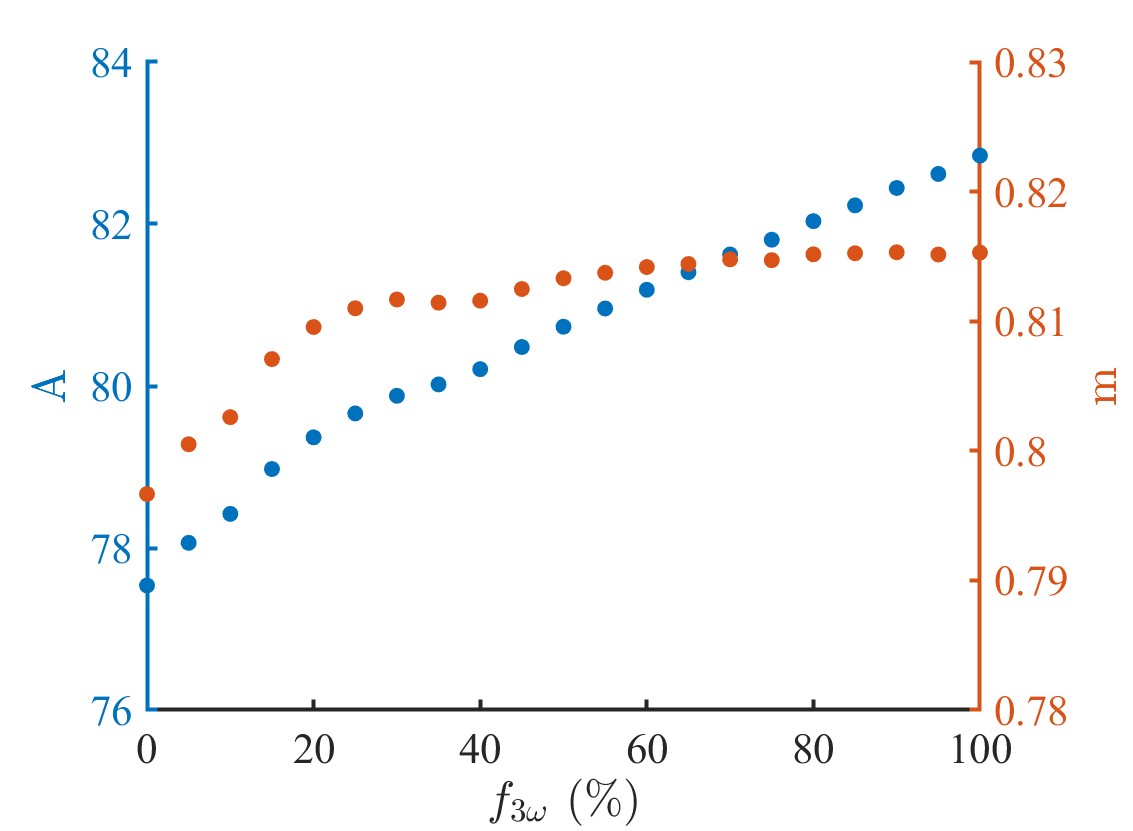}
	\caption{
		Dependence of the steady-ablation fitting parameters $A$ and $m$ on the $3\omega$ fraction $f_{3\omega}$. The ablation pressure is fitted by $P_a=A I^m$ using the pressure values at $t=5~\mathrm{ns}$ for the $600~\mu\mathrm{m}$ CH target. The coefficient $A$ increases monotonically with $f_{3\omega}$, while the exponent $m$ remains in a narrow range of approximately $0.80$--$0.82$.
	}
	\label{fig:A_m_steady}
\end{figure}

The above analysis is repeated for different values of $f_{3\omega}$. Figure~\ref{fig:A_m_steady} shows the fitted parameters $A$ and $m$ as functions of the $3\omega$ fraction. The coefficient $A$ increases monotonically with $f_{3\omega}$, from approximately $77.5$ at $f_{3\omega}=0\%$ to approximately $83$ at $f_{3\omega}=100\%$. This trend indicates that replacing part of the $2\omega$ drive with $3\omega$ light increases the ablation pressure at fixed total laser intensity. In contrast, the exponent $m$ changes only weakly with $f_{3\omega}$. Over the full range from pure $2\omega$ to pure $3\omega$, the fitted exponent remains within a narrow interval,
\begin{equation}
m \simeq 0.80\text{--}0.82 .
\end{equation}
Thus, in the quasi-steady ablation regime, increasing the $3\omega$ fraction mainly modifies the normalization of the pressure scaling rather than the intensity dependence itself. The pressure scaling can therefore be written as
\begin{equation}
P_a(I,f_{3\omega}) = A(f_{3\omega}) I^{m(f_{3\omega})},
\label{eq:Pa_f_scaling}
\end{equation}
with $A(f_{3\omega})$ increasing monotonically and $m(f_{3\omega})$ remaining nearly constant. The increase of $A$ with $f_{3\omega}$ is qualitatively consistent with the known wavelength dependence of the ablation pressure. Since $3\omega$ light has a shorter wavelength than $2\omega$ light, a larger $3\omega$ fraction is expected to increase the ablation pressure under otherwise identical irradiation conditions. However, the mixed-wavelength drive cannot be rigorously represented by a single effective wavelength without additional modeling of absorption, energy deposition, and transport. We therefore do not impose an analytic expression for $A(f_{3\omega})$ or $m(f_{3\omega})$. Instead, the fitted parameters obtained from the FLASH data are used directly in the following analysis.

\subsection{Ablation-pressure scaling during foil acceleration}

\begin{figure}[htbp]
    \centering
    \includegraphics[width=\textwidth]{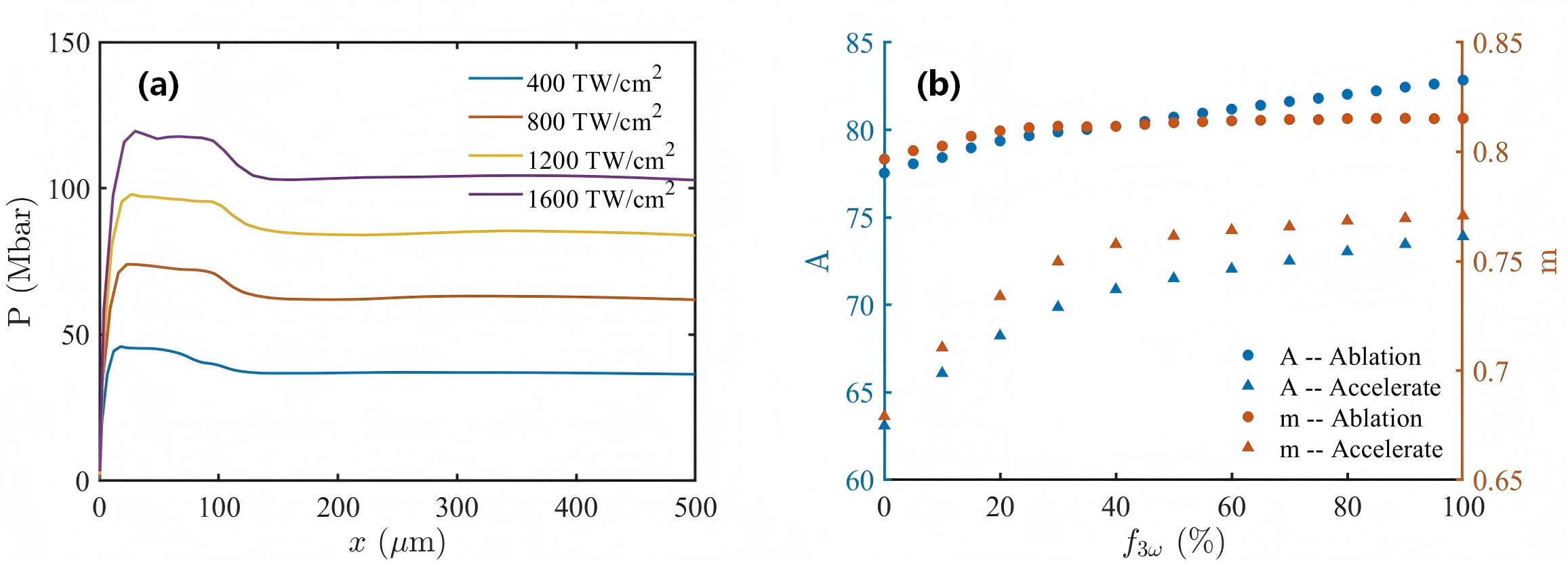}
    \caption{
    Ablation-pressure scaling during the foil-acceleration stage. (a) Ablation pressure as a function of the ablation-front propagation distance $x$ for a $100~\mu\mathrm{m}$ CH target at $f_{3\omega}=50\%$. The pressure first rises rapidly and then decreases after the ablation front has propagated by approximately $100~\mu\mathrm{m}$, which is attributed to the arrival of the rarefaction wave from the rear surface. At $x=500~\mu\mathrm{m}$, the pressure approaches a quasi-steady acceleration-stage value. (b) Comparison of the fitted parameters $A$ and $m$ between the thick-target steady-ablation regime and the foil-acceleration stage. The acceleration-stage scaling exhibits a lower normalization coefficient and a smaller intensity exponent, with $m_{\mathrm{acc}}\simeq0.68$--$0.77$.
    }
    \label{fig:acc_scaling}
\end{figure}

The preceding results characterize the quasi-steady ablation of a thick target. In practical foil acceleration, however, the finite target thickness and rarefaction-wave dynamics can modify the effective ablation pressure. To examine this effect, we perform simulations for a $100~\mu\mathrm{m}$ CH target and analyze the pressure evolution as a function of the ablation-front propagation distance $x$. Figure~\ref{fig:acc_scaling}(a) shows the ablation pressure as a function of $x$ for $f_{3\omega}=50\%$. The pressure rises rapidly at the beginning of the interaction and reaches a high value within the first several tens of micrometers. When the ablation front has propagated by approximately $100~\mu\mathrm{m}$, the pressure starts to decrease. This pressure reduction is attributed to the arrival of the rarefaction wave from the rear surface of the finite-thickness foil. After the rarefaction wave reaches the ablation front, the foil begins to accelerate as a whole, and the local pressure structure is modified. As the ablation front continues to move forward, the pressure gradually approaches a new quasi-steady level. At $x=500~\mu\mathrm{m}$, the pressure has become nearly constant for all laser intensities considered. In this paper, the ablation quantities at $x=600~\mu\mathrm{m}$ are used to characterize the acceleration stage.

The scaling parameters obtained in the acceleration stage are compared with those in the steady-ablation stage in Fig.~\ref{fig:acc_scaling}(b). The normalization coefficient in the acceleration stage is systematically lower than that in the thick-target ablation case. For example, $A$ in the steady-ablation regime lies approximately in the range $77.5$--$83$, whereas the acceleration-stage value lies approximately in the range $63$--$74$. This reduction reflects the effect of finite foil thickness and rarefaction-wave dynamics, which lower the effective ablation pressure during long-distance acceleration. The fitted exponent also changes significantly. In the quasi-steady thick-target ablation regime,
\begin{equation}
m_{\mathrm{abl}} \simeq 0.80\text{--}0.82,
\end{equation}
whereas in the acceleration stage,
\begin{equation}
m_{\mathrm{acc}} \simeq 0.68\text{--}0.77.
\end{equation}
Thus, the intensity dependence of the ablation pressure becomes weaker during foil acceleration. This result shows that the thick-target steady-ablation scaling should not be directly applied to the acceleration stage without accounting for finite-target effects.

Another important feature is that the acceleration-stage parameters are more sensitive to $f_{3\omega}$. Both $A_{\mathrm{acc}}$ and $m_{\mathrm{acc}}$ increase with increasing $f_{3\omega}$. In particular, $m_{\mathrm{acc}}$ rises from approximately $0.68$ for pure $2\omega$ drive to approximately $0.77$ for pure $3\omega$ drive. This behavior indicates that the $3\omega$ component not only increases the pressure amplitude but also modifies the intensity dependence more strongly during foil acceleration than in the thick-target steady-ablation regime.

\subsection{Pressure enhancement due to $3\omega$ mixing}

\begin{figure}[htbp]
    \centering
    \includegraphics[width=\textwidth]{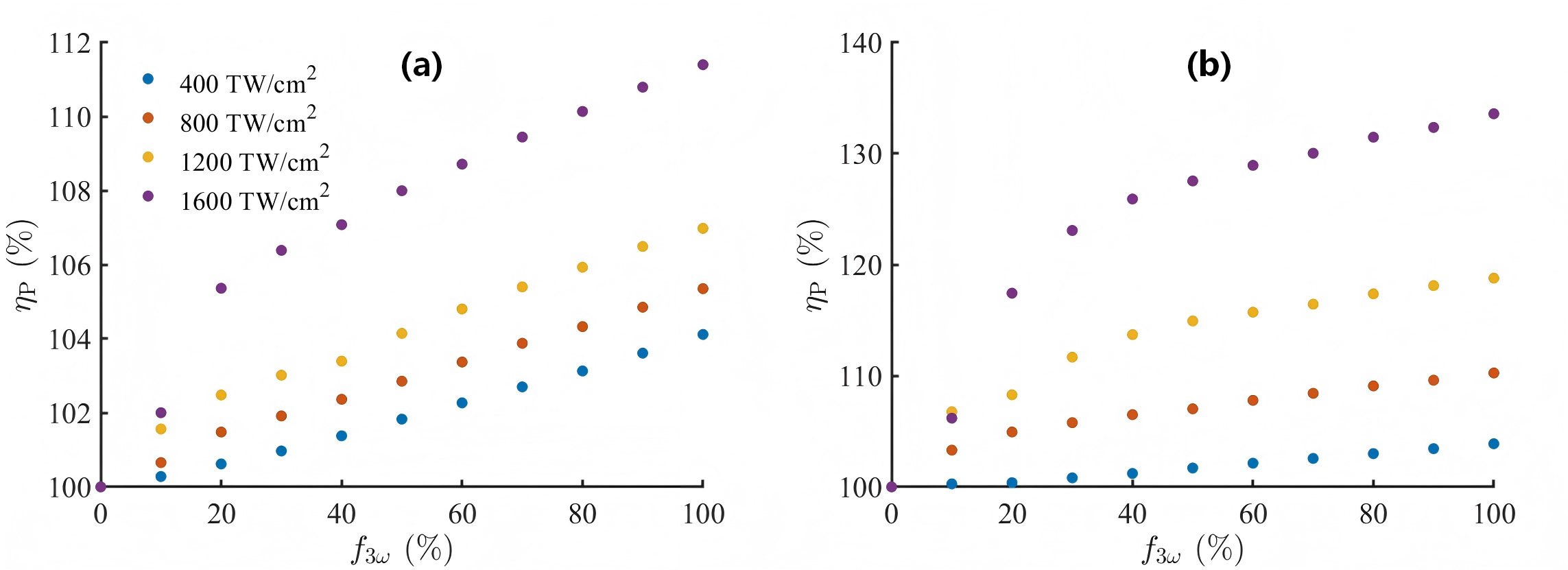}
    \caption{
    Pressure enhancement factor induced by $3\omega$ mixing. The enhancement factor is defined as $\eta_P=P_a(I,f_{3\omega})/P_a(I,0)$, where $P_a(I,0)$ is the ablation pressure under pure $2\omega$ irradiation at the same total laser intensity. (a) Enhancement factor in the thick-target steady-ablation regime. The pressure enhancement is moderate, reaching approximately $110\%$--$112\%$ at high intensity. (b) Enhancement factor in the foil-acceleration stage. The enhancement is significantly stronger during acceleration, reaching approximately $134\%$ for $1600~\mathrm{TW/cm^2}$ at $f_{3\omega}=100\%$.
    }
    \label{fig:etaP}
\end{figure}

To quantify the pressure increase caused by the $3\omega$ component, we define the pressure enhancement factor as
\begin{equation}
\eta_P(I,f_{3\omega})
=
\frac{P_a(I,f_{3\omega})}{P_a(I,0)} ,
\label{eq:etaP}
\end{equation}
where $P_a(I,0)$ is the ablation pressure for pure $2\omega$ irradiation at the same total intensity. Figure~\ref{fig:etaP}(a) shows $\eta_P$ in the quasi-steady ablation regime. For all intensities, $\eta_P$ increases monotonically with $f_{3\omega}$. The enhancement is relatively modest at low intensity and becomes more pronounced at high intensity. At $f_{3\omega}=100\%$, the enhancement is approximately $4\%$ for $400~\mathrm{TW/cm^2}$, approximately $5\%$ for $800~\mathrm{TW/cm^2}$, approximately $7\%$ for $1200~\mathrm{TW/cm^2}$, and about $11\%$--$12\%$ for $1600~\mathrm{TW/cm^2}$. Therefore, in the thick-target steady-ablation regime, replacing $2\omega$ light with $3\omega$ light increases the ablation pressure, but the enhancement remains moderate. Figure~\ref{fig:etaP}(b) shows the corresponding enhancement factor in the acceleration stage. In this case, the enhancement is much stronger, especially at high intensity. At $f_{3\omega}=100\%$, $\eta_P$ is only slightly above unity for $400~\mathrm{TW/cm^2}$, but it increases to approximately $110\%$ for $800~\mathrm{TW/cm^2}$, approximately $119\%$ for $1200~\mathrm{TW/cm^2}$, and approximately $134\%$ for $1600~\mathrm{TW/cm^2}$. This comparison demonstrates that the effect of $3\omega$ mixing is more significant during foil acceleration than during thick-target steady ablation. The stronger enhancement in the acceleration stage is consistent with the larger variation of $A_{\mathrm{acc}}$ and $m_{\mathrm{acc}}$ with $f_{3\omega}$. Since the acceleration dynamics and hydrodynamic instabilities are governed by the pressure during the foil-acceleration stage, the acceleration-stage enhancement factor is the more relevant quantity for evaluating the performance of mixed-wavelength drive schemes.

\section{Linear RTI Gain under a Fixed Final Velocity}
\label{sec:rti_model}

To assess the linear Rayleigh--Taylor instability risk during the acceleration phase, we first construct a simplified theoretical model based on the Takabe formula \cite{Takabe1985,Betti1998}. The purpose of this section is not to reproduce the full hydrodynamic evolution, but to clarify how the acceleration $g$, the ablation velocity $V_a$, and the density-gradient scale length $L_m$ affect the maximum possible RTI amplification when the imploding shell reaches the same final velocity. This analysis provides a theoretical basis for interpreting the FLASH-1D results discussed in the following sections.

The ablative RTI growth rate is written as Eq.~\eqref{eq:takabe_growth}, in which the coefficients $\alpha$ and $\beta$ depend weakly on the Froude number and the density profile. In the following analysis, we use the typical values for a CH ablator \cite{Betti1998},
\begin{equation}
\alpha = 1.0, 
\qquad 
\beta = 1.7 .
\end{equation}
For a constant-acceleration trajectory, the time required for the shell to reach a prescribed final velocity $v_f$ is
\begin{equation}
t_f = \frac{v_f}{g}.
\label{eq:tf}
\end{equation}
In this work, the target final velocity is chosen as
\begin{equation}
v_f = 300~\mathrm{km/s}
=
300~\mu\mathrm{m/ns}.
\end{equation}
The corresponding linear amplification factor for a perturbation mode $k$ can therefore be expressed as
\begin{equation}
G(k,g)
=
\exp
\left[
\gamma(k,g)
\frac{v_f}{g}
\right],
\label{eq:gain_def}
\end{equation}
or equivalently,
\begin{equation}
\ln G(k,g)
=
\left[
\alpha
\sqrt{
\frac{k g}{1+kL_m}
}
-
\beta kV_a
\right]
\frac{v_f}{g}.
\label{eq:ln_gain}
\end{equation}
Because a realistic surface perturbation generally contains a broad spectrum of modes, the most dangerous mode is not fixed a priori. For each acceleration $g$, we define the gain envelope as the maximum gain over all wavenumbers,
\begin{equation}
G_{\mathrm{env}}(g)
=
\max_k G(k,g),
\label{eq:gain_env}
\end{equation}
and the corresponding most dangerous wavenumber as
\begin{equation}
k_{\mathrm{max}}(g)
=
\arg\max_k G(k,g).
\label{eq:kmax_def}
\end{equation}
The envelope $G_{\mathrm{env}}(g)$ represents the largest linear RTI amplification attainable at a given acceleration under the fixed final-velocity constraint.

\subsection{Analytical limit for $L_m=0$}
\label{subsec:Lm_zero}

We first consider the limiting case without density-gradient stabilization, i.e., $L_m=0$. In this case, Eq.~\eqref{eq:takabe_growth} reduces to
\begin{equation}
\gamma(k,g)
=
\alpha \sqrt{kg}
-
\beta kV_a .
\label{eq:takabe_Lm0}
\end{equation}
Since the exponential function is monotonic, maximizing $G(k,g)$ is equivalent to maximizing the growth-rate factor
\begin{equation}
F(k)
=
\alpha \sqrt{kg}
-
\beta kV_a .
\label{eq:Fk_Lm0}
\end{equation}
Taking the derivative of $F(k)$ with respect to $k$, one obtains
\begin{equation}
\frac{dF}{dk}
=
\frac{\alpha\sqrt{g}}{2\sqrt{k}}
-
\beta V_a .
\label{eq:dFdk}
\end{equation}
The extremum condition $dF/dk=0$ gives
\begin{equation}
\sqrt{k^*}
=
\frac{\alpha\sqrt{g}}{2\beta V_a},
\end{equation}
and hence
\begin{equation}
k^*
=
\frac{\alpha^2 g}{4\beta^2 V_a^2}.
\label{eq:kstar_Lm0}
\end{equation}
The second derivative,
\begin{equation}
\frac{d^2F}{dk^2}
=
-
\frac{\alpha\sqrt{g}}{4k^{3/2}}
<0 ,
\end{equation}
confirms that this extremum is a maximum for $k>0$. Substituting Eq.~\eqref{eq:kstar_Lm0} into Eq.~\eqref{eq:Fk_Lm0} yields
\begin{equation}
F(k^*)
=
\frac{\alpha^2 g}{4\beta V_a}.
\label{eq:Fkstar_Lm0}
\end{equation}
Therefore, the logarithm of the maximum gain becomes
\begin{equation}
\ln G_{\mathrm{env}}
=
F(k^*)\frac{v_f}{g}
=
\frac{\alpha^2 v_f}{4\beta V_a}.
\label{eq:lnGenv_Lm0}
\end{equation}
For $v_f=300~\mu\mathrm{m/ns}$, this gives
\begin{equation}
G_{\mathrm{env}}
=
\exp
\left(
\frac{75\alpha^2}{\beta V_a}
\right).
\label{eq:Genv_Lm0_300}
\end{equation}

\begin{figure}[htbp]
    \centering
    \includegraphics[width=\textwidth]{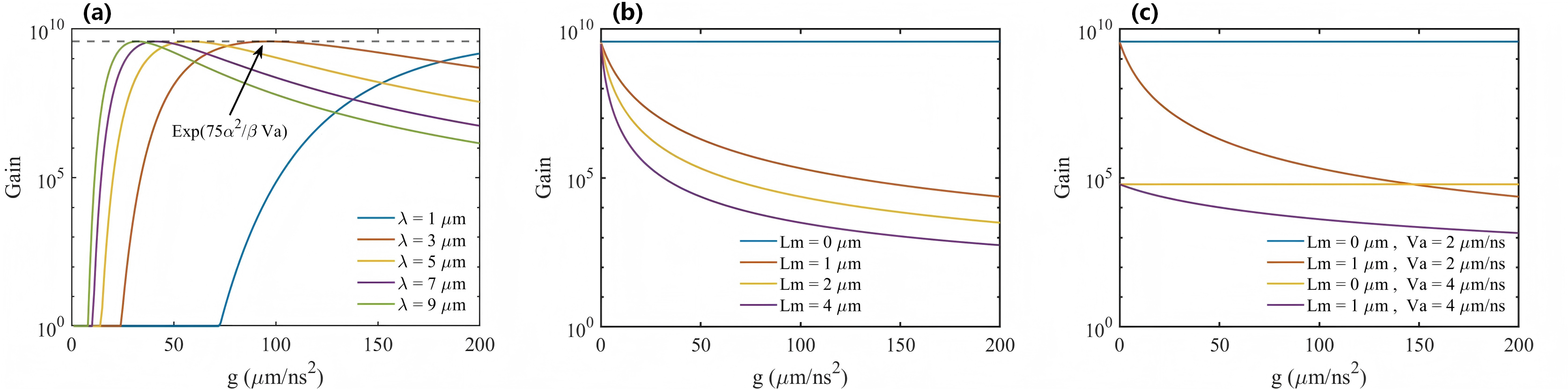}
    \caption{
    Linear RTI gain envelope under a fixed final velocity of $v_f=300~\mathrm{km/s}$. (a) Gain curves of different perturbation modes for $L_m=0$ and $V_a=2~\mu\mathrm{m/ns}$. The upper envelope agrees with the analytical prediction $G_{\mathrm{env}}=\exp(75\alpha^2/\beta V_a)$. (b) Effect of the density-gradient scale length $L_m$ on the gain envelope at fixed $V_a=2~\mu\mathrm{m/ns}$. (c) Comparison of gain envelopes for different combinations of $V_a$ and $L_m$. Increasing either $V_a$ or $L_m$ reduces the maximum RTI gain.
    }
    \label{fig:rti_envelope}
\end{figure}

Equation~\eqref{eq:Genv_Lm0_300} is an important result: in the $L_m=0$ limit, the envelope of the most dangerous RTI gain is independent of the acceleration $g$ when the final velocity is fixed. Although a larger $g$ increases the instantaneous RTI growth rate, it simultaneously shortens the acceleration duration $t_f=v_f/g$. For the most dangerous mode, these two effects exactly compensate each other. Figure~\ref{fig:rti_envelope}(a) illustrates this analytical behavior. Individual modes with different wavelengths reach their largest gains at different accelerations. However, their upper envelope remains approximately constant and agrees with the analytical prediction in Eq.~\eqref{eq:Genv_Lm0_300}. This result demonstrates that, without density-gradient stabilization, modifying the acceleration alone does not reduce the maximum possible RTI gain under a fixed final-velocity constraint.

\subsection{Numerical gain envelope for finite $L_m$}
\label{subsec:finite_Lm}

We next include the finite density-gradient scale length $L_m$. The growth rate is then described by Eq.~\eqref{eq:takabe_growth}, and the gain is evaluated from Eq.~\eqref{eq:gain_def}. In contrast to the $L_m=0$ case, the analytical maximization over $k$ becomes more complicated because the stabilizing factor $(1+kL_m)^{-1/2}$ modifies the $k$-dependence of the RTI drive. Therefore, the envelope $G_{\mathrm{env}}(g)$ is obtained numerically by maximizing $G(k,g)$ over $k$ at each acceleration.

Figure~\ref{fig:rti_envelope}(b) shows the calculated gain envelope for $V_a=2~\mu\mathrm{m/ns}$ with different values of $L_m$. The case $L_m=0$ recovers the nearly constant envelope predicted by Eq.~\eqref{eq:Genv_Lm0_300}. Once a finite density-gradient scale length is introduced, the envelope decreases with increasing acceleration. In addition, a larger $L_m$ leads to a lower gain envelope over the entire acceleration range. This behavior can be understood from the RTI drive term in Eq.~\eqref{eq:takabe_growth},
\begin{equation}
\sqrt{
\frac{k g}{1+kL_m}
}.
\label{eq:rt_drive_finiteLm}
\end{equation}
For $L_m>0$, the denominator $1+kL_m$ weakens the effective RTI drive, especially for short-wavelength modes with large $k$. Therefore, a finite density-gradient scale length suppresses high-$k$ perturbations and reduces the maximum attainable gain. The exact cancellation between the increased instantaneous growth rate and the reduced acceleration time, which occurs in the $L_m=0$ limit, no longer holds when $L_m$ is finite.

Figure~\ref{fig:rti_envelope}(c) further compares the effects of $V_a$ and $L_m$. Increasing the ablation velocity from $V_a=2~\mu\mathrm{m/ns}$ to $V_a=4~\mu\mathrm{m/ns}$ substantially reduces the gain envelope. This is due to the ablative stabilization term,
\begin{equation}
-\beta kV_a ,
\end{equation}
which becomes stronger for larger $V_a$. The combined increase of $V_a$ and $L_m$ produces the strongest reduction of the gain envelope. These results indicate that, for implosions reaching the same final velocity, the RTI is more effectively mitigated by increasing the ablation velocity and density-gradient scale length. However, under the condition of a finite-$L_m$, increasing $g$ also tends to be beneficial for controlling RTI.

\subsection{Peak-point analysis for fixed-$k$ modes}
\label{subsec:peak_point}

To further clarify the behavior of the numerical envelope for finite $L_m$, we analyze the gain of each fixed-$k$ mode as a function of acceleration. From Eq.~\eqref{eq:ln_gain}, the logarithmic gain can be rewritten as
\begin{equation}
\ln G(k,g)
=
v_f
\left[
A g^{-1/2}
-
B g^{-1}
\right],
\label{eq:lnG_AB}
\end{equation}
where
\begin{equation}
A
=
\alpha
\sqrt{
\frac{k}{1+kL_m}
},
\qquad
B
=
\beta kV_a .
\label{eq:AB_def}
\end{equation}
For a fixed mode $k$, the extremum of $\ln G(k,g)$ with respect to $g$ is determined by
\begin{equation}
\frac{d}{dg}
\left[
A g^{-1/2}
-
B g^{-1}
\right]
=
-\frac{A}{2}g^{-3/2}
+
Bg^{-2}
=
0 .
\label{eq:dgd_lnG}
\end{equation}
This gives
\begin{equation}
\sqrt{g^*}
=
\frac{2B}{A},
\end{equation}
or
\begin{equation}
g^*(k)
=
\frac{4B^2}{A^2}.
\label{eq:gstar_AB}
\end{equation}
Substituting the definitions of $A$ and $B$, we obtain
\begin{equation}
g^*(k)
=
\frac{
4\beta^2 k V_a^2(1+kL_m)
}
{\alpha^2}.
\label{eq:gstar_k}
\end{equation}
The corresponding maximum value of the logarithmic gain is
\begin{equation}
\ln G^*(k)
=
v_f
\frac{A^2}{4B}.
\label{eq:lnGstar_AB}
\end{equation}
Expanding $A$ and $B$ gives
\begin{equation}
\ln G^*(k)
=
\frac{
v_f \alpha^2
}
{
4\beta V_a(1+kL_m)
}.
\label{eq:lnGstar_k}
\end{equation}
Therefore,
\begin{equation}
G^*(k)
=
\exp
\left[
\frac{
v_f \alpha^2
}
{
4\beta V_a(1+kL_m)
}
\right].
\label{eq:Gstar_k}
\end{equation}

\begin{figure}[htbp]
    \centering
    \includegraphics[width=\textwidth]{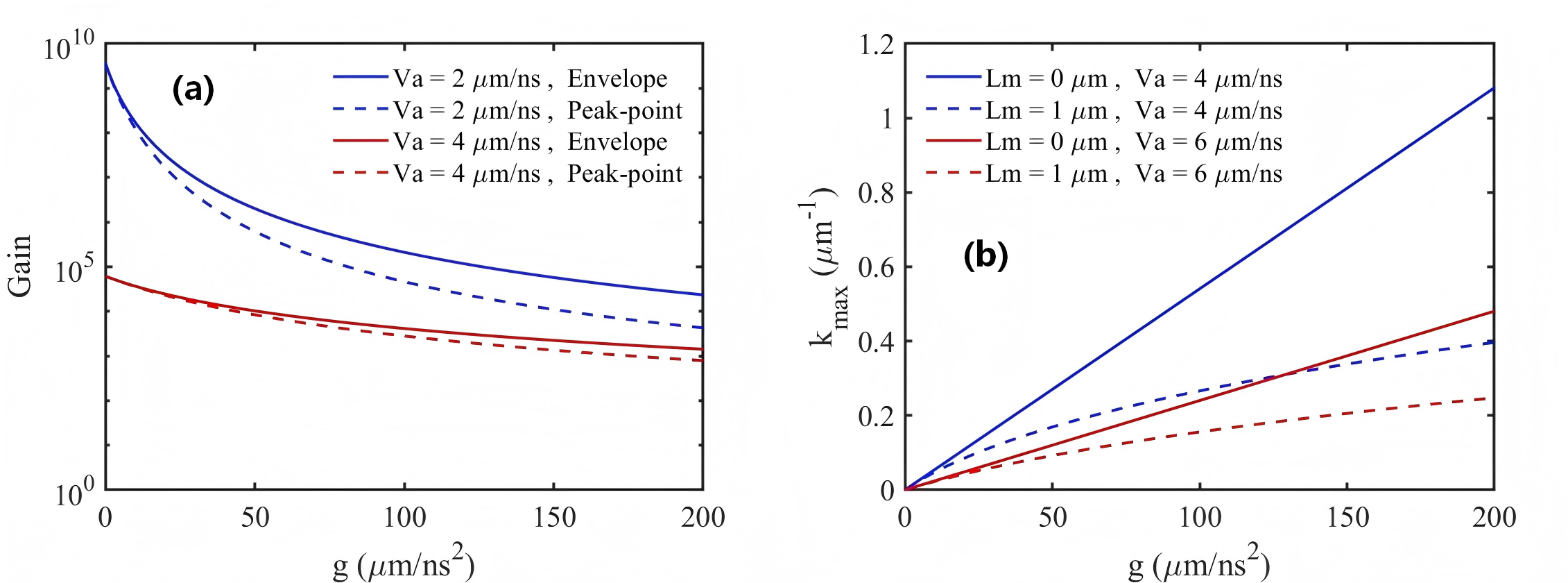}
    \caption{
    Peak-point analysis and evolution of the most dangerous mode. (a) Comparison between the numerical gain envelope and the locus of the individual peak points $(g^*(k),G^*(k))$ for different ablation velocities. The peak-point curve captures the overall trend of the numerical envelope. (b) Most dangerous wavenumber $k_{\mathrm{max}}$ as a function of acceleration $g$. Increasing $V_a$ or $L_m$ shifts the most dangerous mode toward smaller $k$, corresponding to longer wavelengths.
    }
    \label{fig:peak_kmax}
\end{figure}

Equation~\eqref{eq:lnGstar_k} shows explicitly that increasing $V_a$ reduces the maximum gain of every fixed-$k$ mode. A finite $L_m$ further suppresses the gain, and this suppression becomes stronger for larger $k$. Therefore, the density-gradient scale length mainly stabilizes short-wavelength modes. In the special case $L_m=0$, Eq.~\eqref{eq:lnGstar_k} reduces to
\begin{equation}
\ln G^*
=
\frac{v_f\alpha^2}{4\beta V_a},
\end{equation}
which is independent of $k$ and consistent with the analytical envelope in Eq.~\eqref{eq:lnGenv_Lm0}. Thus, the finite-$L_m$ result can be regarded as a generalization of the $L_m=0$ analytical limit. Figure~\ref{fig:peak_kmax}(a) compares the numerical envelope $G_{\mathrm{env}}(g)$ with the locus of the individual peak points $(g^*(k),G^*(k))$. Although the peak-point curve does not exactly coincide with the numerical envelope, it captures its overall decreasing trend and its dependence on $V_a$. This comparison suggests that the peak-point analysis provides a useful semi-analytical interpretation of the finite-$L_m$ envelope.

In addition to the magnitude of the gain envelope, the corresponding most dangerous wavenumber $k_{\mathrm{max}}$ is also important. The mode $k_{\mathrm{max}}$ determines which part of the perturbation spectrum is most amplified during the acceleration phase. Numerically, $k_{\mathrm{max}}(g)$ is obtained from Eq.~\eqref{eq:kmax_def}. Figure~\ref{fig:peak_kmax}(b) shows the $k_{\mathrm{max}}$ as a function of acceleration for different values of $V_a$ and $L_m$. In general, $k_{\mathrm{max}}$ increases with $g$. This trend is consistent with the analytical $L_m=0$ result,
\begin{equation}
k^* = \frac{\alpha^2 g}{4\beta^2V_a^2},
\end{equation}
which predicts a linear increase of the most dangerous wavenumber with acceleration. Figure~\ref{fig:peak_kmax}(b) also shows that increasing $V_a$ lowers $k_{\mathrm{max}}$. This is because stronger ablative stabilization penalizes large-$k$ modes through the term $-\beta kV_a$. Similarly, increasing $L_m$ also lowers $k_{\mathrm{max}}$, especially at larger $g$. Equivalently, the most dangerous wavelength,
\begin{equation}
\lambda_{\mathrm{max}} = \frac{2\pi}{k_{\mathrm{max}}},
\end{equation}
shifts toward longer wavelengths when $V_a$ or $L_m$ is increased. Thus, larger ablation velocity and density-gradient scale length not only reduce the overall RTI gain, but also modify the spectral location of the most dangerous mode. This is particularly relevant for direct-drive implosions, where surface roughness and laser-imprint perturbations can contain significant short- and intermediate-wavelength components.

\section{RTI Mitigation during Foil Acceleration}
\label{sec:rti_results}

The gain-envelope analysis in Sec.~\ref{sec:rti_model} shows that, under a fixed final-velocity constraint, the Rayleigh--Taylor instability is not determined by the shell acceleration alone. Instead, the ablation velocity $V_a$ and the density-gradient scale length $L_m$ play essential roles in reducing the maximum linear amplification. In this section, we apply this framework to the hydrodynamic quantities extracted from the FLASH simulations of accelerating CH foils and evaluate the effect of mixed $2\omega$--$3\omega$ irradiation on the RTI gain.

\subsection{Reduction of the maximum RTI gain}
\label{sec:rti_results_A}

\begin{figure}[htbp]
\centering
\includegraphics[width=\linewidth]{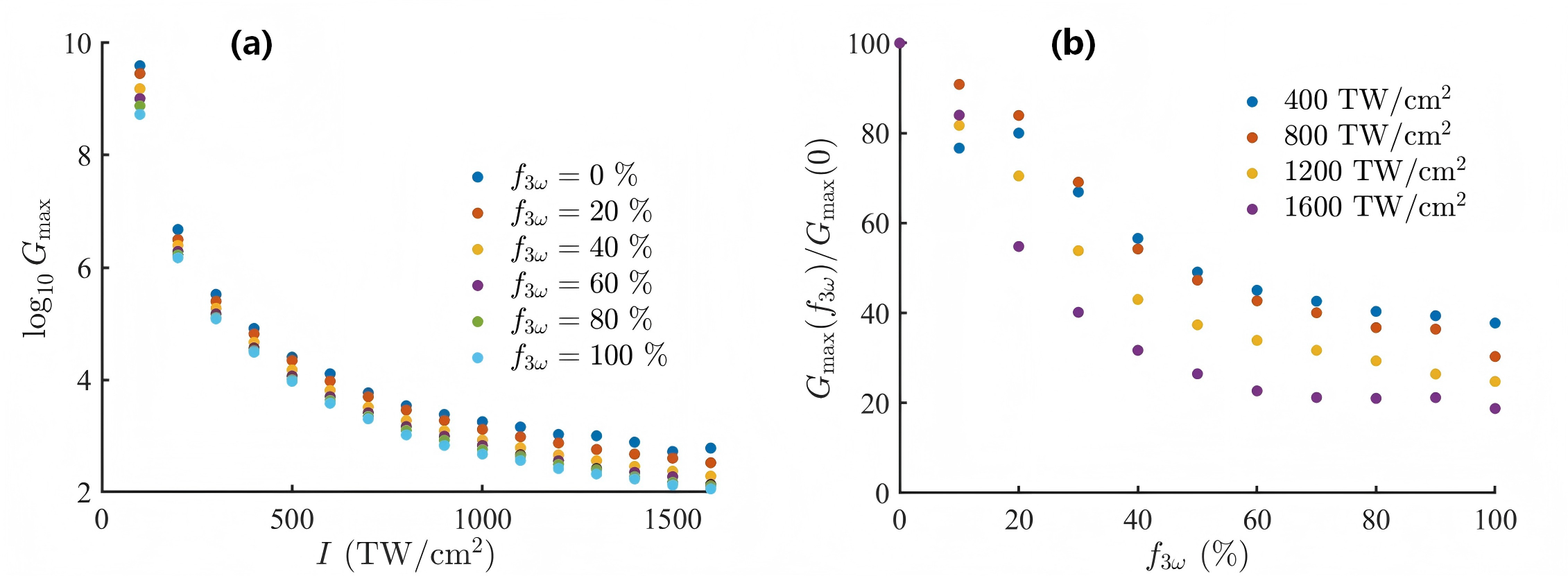}
\caption{
Maximum linear RTI gain during the acceleration of a $100~\mu\mathrm{m}$ CH foil. (a) $\log_{10}G_{\max}$ as a function of the total incident laser intensity for different $3\omega$ fractions. The gain decreases rapidly with increasing intensity. (b) Normalized gain $G_{\max}(f_{3\omega})/G_{\max}(0)$ as a function of $f_{3\omega}$ for several laser intensities. Increasing the $3\omega$ fraction reduces the maximum RTI gain, and the reduction is strongest at high intensity.
}
\label{fig:rti_gain}
\end{figure}

Figure~\ref{fig:rti_gain}(a) shows the maximum RTI gain as a function of the total laser intensity for different $3\omega$ fractions. The gain is plotted in terms of $\log_{10}G_{\max}$. At low intensity, the linear amplification is large. For $I\simeq 100~\mathrm{TW/cm^2}$, $\log_{10}G_{\max}$ is of the order of $9$--$10$, indicating strong RTI amplification during the relatively long and weak acceleration process. As the laser intensity increases, $\log_{10}G_{\max}$ decreases rapidly. At $I\simeq 1600~\mathrm{TW/cm^2}$, the maximum gain is reduced to $\log_{10}G_{\max}\simeq 2$--$3$, depending on the value of $f_{3\omega}$. This strong reduction with increasing intensity reflects the combined effect of a shorter acceleration time, a larger ablation velocity, and a modified density-gradient scale length. Although a larger acceleration $g$ tends to increase the instantaneous RTI drive, the acceleration duration is reduced at the same time. More importantly, the higher-intensity cases also exhibit larger $V_a$, which enhances the ablative stabilization term $-\beta k V_a$ in Eq.~\eqref{eq:takabe_growth}. Therefore, the maximum gain decreases monotonically as the drive intensity is increased.

The effect of the $3\omega$ fraction is highlighted in Fig.~\ref{fig:rti_gain}(b), where the gain is normalized to the pure $2\omega$ result at the same total laser intensity,
\begin{equation}
R_G(I,f_{3\omega})=
\frac{G_{\max}(I,f_{3\omega})}
     {G_{\max}(I,0)} .
\label{eq:gain_reduction_factor}
\end{equation}
For all intensities considered here, $R_G$ decreases as $f_{3\omega}$ increases. This indicates that replacing part of the $2\omega$ drive by $3\omega$ light reduces the maximum linear RTI amplification. The reduction is more pronounced at higher laser intensity. For example, at $I=1600~\mathrm{TW/cm^2}$, increasing $f_{3\omega}$ from 0 to $100\%$ reduces $G_{\max}$ to approximately $20\%$ of the pure $2\omega$ value. At lower intensity, the reduction is weaker but remains systematic. Thus, $3\omega$ mixing provides an additional stabilizing effect during foil acceleration.

\subsection{Hydrodynamic origin of the gain reduction}
\label{sec:rti_results_B}

\begin{figure}[htbp]
\centering
\includegraphics[width=\linewidth]{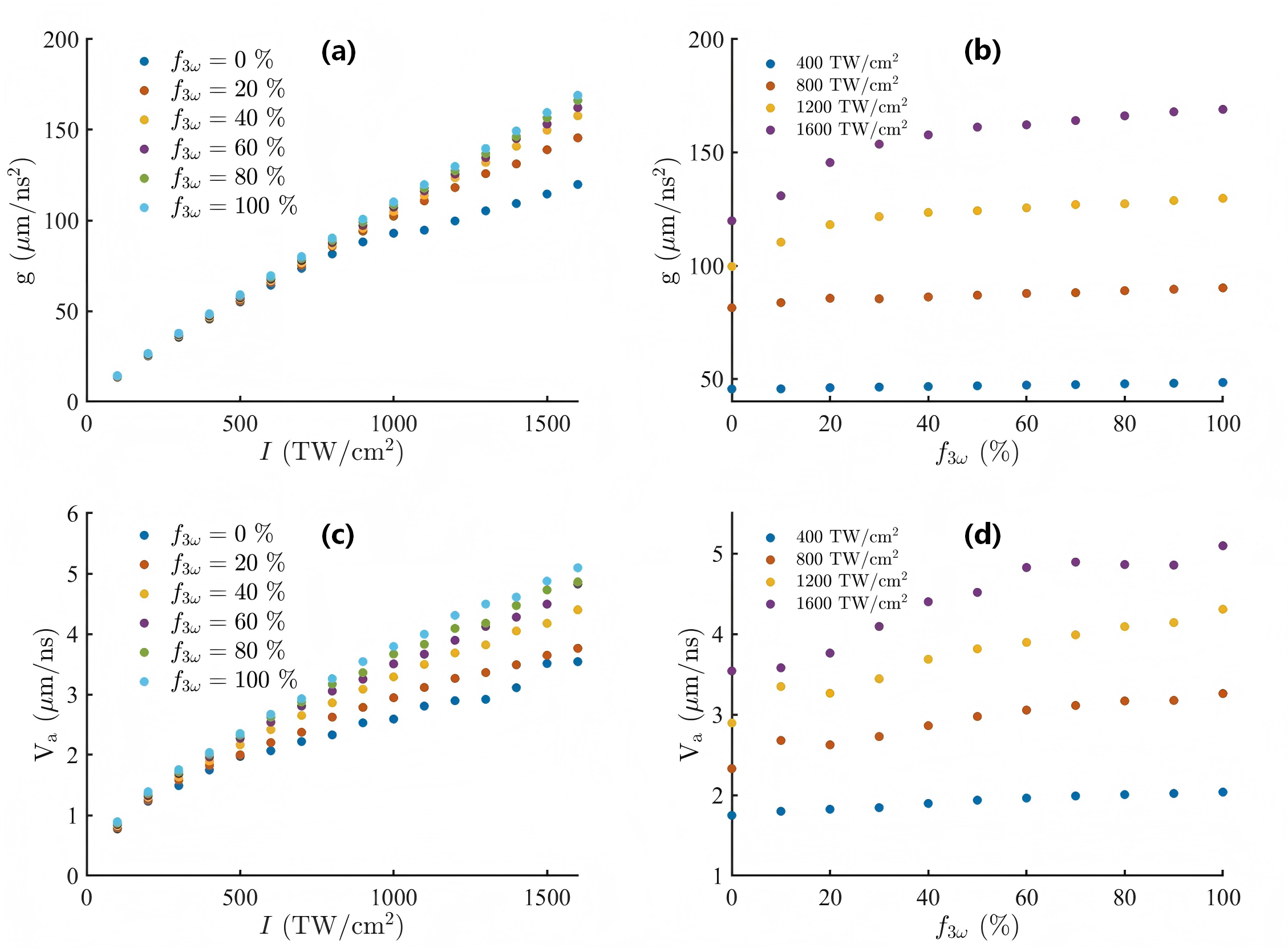}
\caption{
Hydrodynamic parameters extracted from the acceleration stage of the $100~\mu\mathrm{m}$ CH foil. (a) Shell acceleration $g$ as a function of the total laser intensity for different $3\omega$ fractions. (b) Shell acceleration as a function of $f_{3\omega}$ for selected intensities. (c) Ablation velocity $V_a$ as a function of intensity. (d) Ablation velocity as a function of $f_{3\omega}$.
}
\label{fig:g_va}
\end{figure}

To identify the hydrodynamic mechanism responsible for the gain reduction, we examine the acceleration $g$, the ablation velocity $V_a$, and the density-gradient scale length $L_m$ extracted from the simulations. Figure~\ref{fig:g_va}(a) shows the shell acceleration as a function of the total laser intensity. The acceleration increases nearly monotonically with intensity for all mixing ratios. At fixed intensity, a larger $f_{3\omega}$ also produces a larger acceleration, especially in the high-intensity regime. This behavior is consistent with the acceleration-stage pressure scaling discussed in Sec.~\ref{sec:pressure}: increasing the $3\omega$ fraction enhances the effective ablation pressure and therefore increases the shell acceleration.
Figure~\ref{fig:g_va}(b) shows the same acceleration data as a function of $f_{3\omega}$ for selected laser intensities. At $I=400~\mathrm{TW/cm^2}$, the acceleration changes only weakly with the mixing ratio. In contrast, for $1200$ and $1600~\mathrm{TW/cm^2}$, the acceleration increases significantly as $f_{3\omega}$ is increased. As shown in Sec.~\ref{sec:rti_model}, under conditions of finite $L_m$, increasing the acceleration $g$ can reduce the maximum RTI gain. Therefore, the results shown above indicate that increasing either the laser intensity $I$ or the mixing fraction of the $3\omega$ laser, $f_{3\omega}$, can effectively suppress the risk of RTI by enhancing $g$. Furthermore, the ablation velocity is a more direct stabilizing parameter. Figure~\ref{fig:g_va}(c) shows that $V_a$ increases systematically with the total laser intensity. For the present cases, $V_a$ increases from approximately $1~\mu\mathrm{m/ns}$ at low intensity to $3$--$5~\mu\mathrm{m/ns}$ at the highest intensities. At fixed intensity, $V_a$ also increases with $f_{3\omega}$, as shown in Fig.~\ref{fig:g_va}(d). The increase is particularly strong at high intensity: for $I=1600~\mathrm{TW/cm^2}$, $V_a$ increases from about $3.6~\mu\mathrm{m/ns}$ for pure $2\omega$ drive to about $5~\mu\mathrm{m/ns}$ for pure $3\omega$ drive.

\begin{figure}[htbp]
\centering
\includegraphics[width=\linewidth]{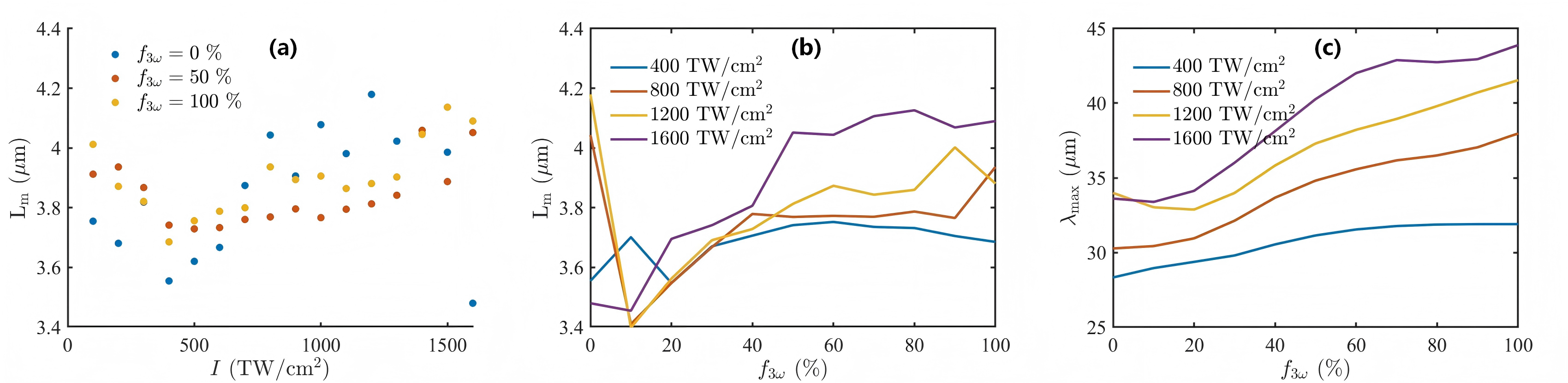}
\caption{
Density-gradient scale length and most dangerous wavelength during the foil acceleration. (a) $L_m$ as a function of the total laser intensity for representative $3\omega$ fractions. (b) $L_m$ as a function of $f_{3\omega}$ for selected intensities. (c) Most amplified wavelength $\lambda_{\max}=2\pi/k_{\max}$ as a function of $f_{3\omega}$. Increasing $f_{3\omega}$ shifts the most dangerous mode toward longer wavelengths.
}
\label{fig:lm_lambda}
\end{figure}

The density-gradient scale length $L_m$ provides an additional stabilization mechanism by reducing the effective RTI drive at large wavenumber. Figure~\ref{fig:lm_lambda}(a) shows $L_m$ as a function of laser intensity for representative mixing ratios. The values of $L_m$ remain in the range of approximately $3.4$--$4.2~\mu\mathrm{m}$. Compared with the monotonic trends of $g$ and $V_a$, the dependence of $L_m$ on intensity is weaker and contains more scatter. Nevertheless, a general increase of $L_m$ with $f_{3\omega}$ can be observed in Fig.~\ref{fig:lm_lambda}(b), especially at higher intensities. For perturbations with large $k$, the factor $1+kL_m$ reduces the effective RTI drive. Therefore, an increase in $L_m$ preferentially suppresses short-wavelength modes. Although the variation of $L_m$ is modest compared with that of $V_a$, it contributes to the reduction of $G_{\max}$ and modifies the spectral location of the most unstable mode. Figure~\ref{fig:lm_lambda}(c) shows the most amplified wavelength $\lambda_{\max}$ as a function of $f_{3\omega}$. The value of $\lambda_{\max}$ lies in the range of approximately $28$--$44~\mu\mathrm{m}$ for the present drive conditions. For a fixed laser intensity, $\lambda_{\max}$ generally increases with $f_{3\omega}$. The shift is more pronounced at higher intensity. For example, at $I=1600~\mathrm{TW/cm^2}$, $\lambda_{\max}$ increases from about $33~\mu\mathrm{m}$ for pure $2\omega$ drive to more than $40~\mu\mathrm{m}$ as the $3\omega$ fraction approaches unity.

\subsection{Implications for mixed-wavelength drive design}

The results in Secs.~\ref{sec:rti_results_A} and \ref{sec:rti_results_B} show that increasing the laser intensity $I$ and the $3\omega$ fraction $f_{3\omega}$ both reduce the linear maximum RTI gain. This trend is mainly attributed to the increases in the shell acceleration $g$ and the ablation velocity $V_a$, rather than to a significant change in the density-gradient scale length $L_m$. In the simulations, both $g$ and $V_a$ increase markedly with $I$ and $f_{3\omega}$, whereas $L_m$ remains nearly unchanged, typically lying in the range of $3.4\text{--}4.2~\mu\mathrm{m}$. This observation is consistent with the theoretical analysis in Sec.~\ref{sec:rti_model}. For the idealized case of $L_m=0$, increasing $g$ does not modify the envelope of the maximum RTI gain. In contrast, for finite $L_m$, a larger $g$ can significantly reduce this gain envelope. The hydrodynamic profiles obtained in Sec.~\ref{sec:rti_results_B} correspond to the latter regime, since $L_m$ remains finite at approximately $3\text{--}4~\mu\mathrm{m}$. Therefore, the increase in $g$ can effectively contribute to the reduction of the maximum RTI gain. Meanwhile, the increase in $V_a$ provides a direct stabilizing mechanism through the ablative stabilization term $-\beta k V_a$ in the linear growth-rate formula~\eqref{eq:takabe_growth}. The implication for mixed-wavelength drive design is therefore straightforward. Provided that the laser energy can be efficiently absorbed and converted into target ablation, a higher laser intensity and a larger $f_{3\omega}$ are favorable for reducing the linear RTI risk during the acceleration phase. The enhanced $g$ lowers the gain envelope under finite-$L_m$ conditions, while the enhanced $V_a$ directly reduces the RTI growth rate through ablative stabilization. In practical high-intensity direct-drive conditions, LPI may affect laser absorption and energy coupling, and may therefore limit this favorable trend. However, such effects are beyond the scope of the present work.

\section{Laser-Energy requirement}
\label{sec:energy}

The results in Secs.~\ref{sec:pressure}--\ref{sec:rti_results} show that increasing the $3\omega$ fraction enhances the acceleration-stage ablation pressure, increases the ablation velocity, and reduces the linear RTI gain. These trends suggest that mixed $2\omega$--$3\omega$ irradiation can improve both the acceleration performance and the hydrodynamic stability. Here, we further examine the laser-energy requirement for reaching a prescribed target velocity and identify the underlying energy-transport mechanism responsible for the improved drive efficiency.

\begin{figure}[htbp]
\centering
\includegraphics[width=1.00\textwidth]{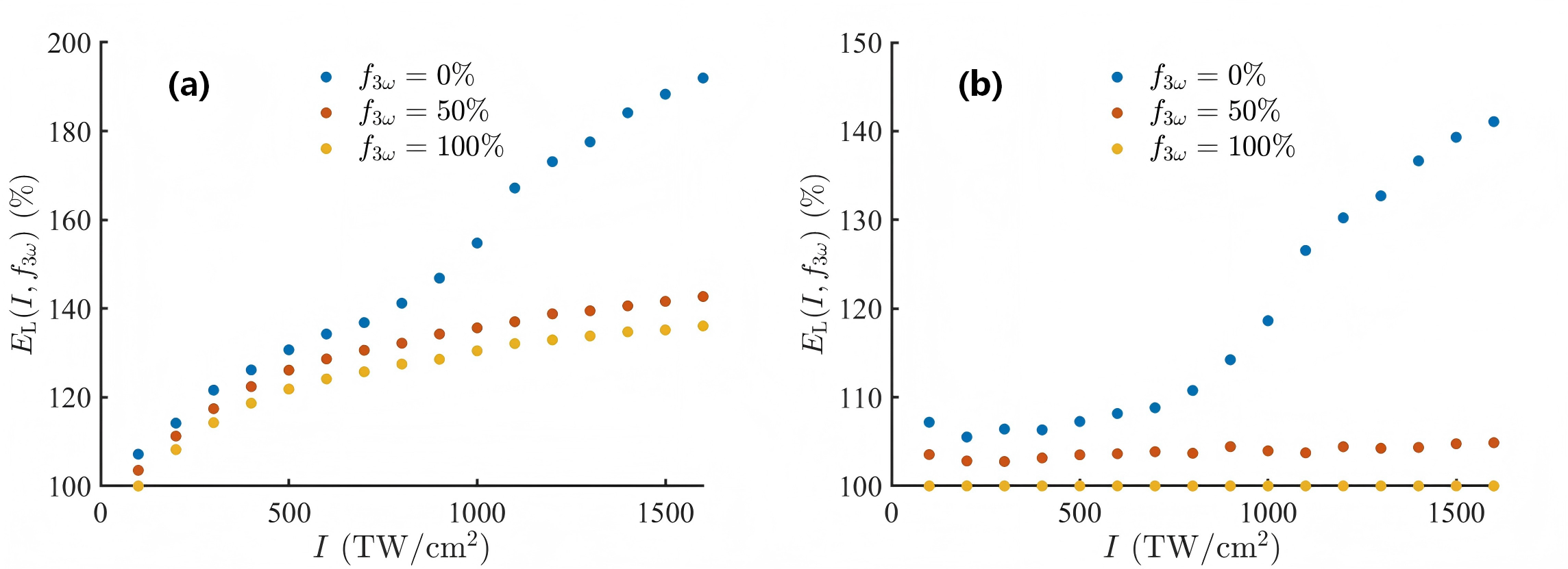}
\caption{
Total incident laser energy required to accelerate the target to $300~{\rm km/s}$ for different $3\omega$ fractions. (a) Laser energy normalized to a fixed reference case. (b) Laser energy normalized to the pure $3\omega$ result at the same total intensity, $E_{\rm L}(I,f_{3\omega})/E_{\rm L}(I,100\%)$. The pure $2\omega$ drive requires the largest laser energy, especially at high intensity, whereas the mixed drive with $f_{3\omega}=50\%$ remains close to the pure $3\omega$ result.
}
\label{fig:energy_requirement}
\end{figure}

\subsection{Laser energy required to reach $300~{\rm km/s}$}

Figure~\ref{fig:energy_requirement} shows the total laser energy required to accelerate the target to $300~{\rm km/s}$. In Fig.~\ref{fig:energy_requirement}(a), the energy is normalized to a fixed reference case $E_{\rm L}(100~\mathrm{TW/cm^2}, 100\%)$. In Fig.~\ref{fig:energy_requirement}(b), the same quantity is normalized to the pure $3\omega$ result at the same total laser intensity,
\begin{equation}
R_E(I,f_{3\omega}) = \frac{E_{\rm L}(I,f_{3\omega})}{E_{\rm L}(I,f_{3\omega}=100\%)} ,
\end{equation}
where $E_{\rm L}$ denotes the incident laser energy required to reach $v_f=300~{\rm km/s}$. Several features can be identified from Fig.~\ref{fig:energy_requirement}. First, for a given laser intensity, the pure $2\omega$ drive requires the largest laser energy, while the pure $3\omega$ drive gives the lowest energy requirement. The equal-intensity mixed drive, $f_{3\omega}=50\%$, lies between these two limits but remains much closer to the pure $3\omega$ case than to the pure $2\omega$ case. In Fig.~\ref{fig:energy_requirement}(b), the energy required by the $f_{3\omega}=50\%$ case is only a few percent larger than that required by the pure $3\omega$ drive over most of the intensity range. Second, the difference between pure $2\omega$ drive and the drives containing a $3\omega$ component becomes more pronounced at high intensity. At $I\simeq 1600~{\rm TW/cm^2}$, the pure $2\omega$ drive requires approximately $40\%$ more laser energy than the pure $3\omega$ drive to reach the same target velocity. In contrast, the $f_{3\omega}=50\%$ case remains close $(\sim 3-5\%)$ to the pure $3\omega$ result. This indicates that adding even a partial $3\omega$ component can recover most of the hydrodynamic efficiency of pure $3\omega$ irradiation. The reduction of the required laser energy is consistent with the acceleration-stage pressure enhancement discussed in Sec.~\ref{sec:pressure}. At fixed total incident intensity, increasing $f_{3\omega}$ increases the effective ablation pressure and the shell acceleration. Therefore, the acceleration time required to reach $300~{\rm km/s}$ is reduced. Since the total incident laser energy is approximately proportional to the product of the incident intensity and the acceleration duration, $E_{\rm L} \propto I \Delta t$, a shorter acceleration time directly lowers the laser energy required to reach the prescribed velocity.

\subsection{Critical-layer location and electron-temperature structure}

\begin{figure}[htbp]
\centering
\includegraphics[width=0.60\textwidth]{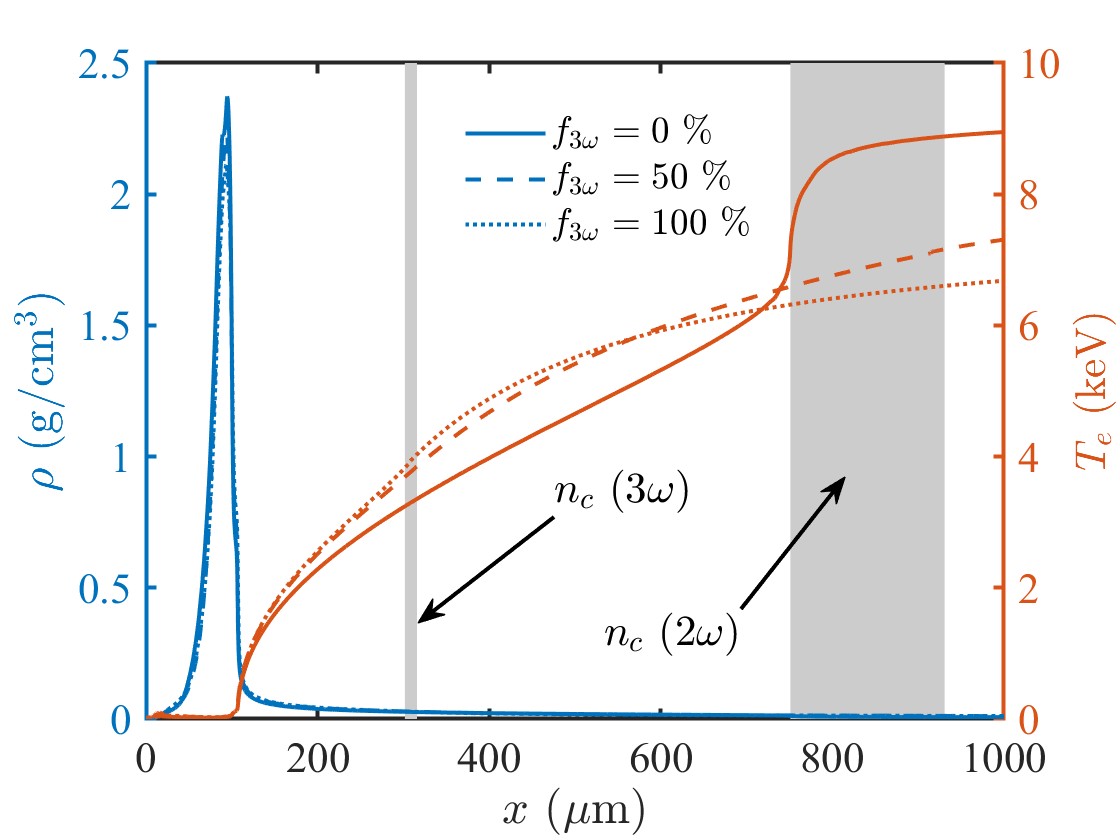}
\caption{
Density and electron-temperature profiles for different $3\omega$ fractions at high laser intensity. The shaded regions indicate the locations of the $3\omega$ and $2\omega$ critical layers. Because $n_c\propto \omega^2$, the $3\omega$ critical layer is located closer to the dense plasma and the ablation region than the $2\omega$ critical layer. Pure $2\omega$ drive produces a hotter outer corona, while the mixed and pure $3\omega$ drives maintain a higher electron temperature near the $3\omega$ critical layer.
}
\label{fig:Te_profile}
\end{figure}

The physical origin of the improved energy utilization can be understood from the different critical densities of the $2\omega$ and $3\omega$ components. Figure~\ref{fig:Te_profile} shows the density and electron-temperature profiles for different values of $f_{3\omega}$ at high laser intensity. The shaded regions indicate the locations of the $3\omega$ and $2\omega$ critical layers. The $3\omega$ critical layer is located much closer to the dense plasma and the ablation region, while the $2\omega$ critical layer is located farther out in the expanding corona. An important feature of Fig.~\ref{fig:Te_profile} is that pure $2\omega$ drive produces the highest electron temperature in the outer coronal plasma. In the region $x\simeq 750$--$930~\mu{\rm m}$, the electron temperature for pure $2\omega$ drive approaches $9~{\rm keV}$, which is higher than the corresponding values for the mixed and pure $3\omega$ drives. This shows that the reduced efficiency of pure $2\omega$ drive is not caused by insufficient coronal heating. Instead, a substantial fraction of the thermal energy is deposited and stored in a low-density region far from the ablation front. Because of the long conduction path and the low density of the outer corona, this thermal energy is not efficiently converted into ablation pressure.

By contrast, the mixed and pure $3\omega$ drives maintain a higher electron temperature near the $3\omega$ critical layer. This region is closer to the dense plasma and to the conduction zone adjacent to the ablation front. Therefore, the heat deposited by the $3\omega$ component is located in a region where it can more effectively contribute to the conductive energy flux toward the ablation front. In this sense, the role of the $3\omega$ component is not merely to increase the overall coronal temperature, but to optimize the spatial distribution of thermal energy. This distinction is essential for understanding the energy-efficiency improvement shown in Fig.~\ref{fig:energy_requirement}. A high electron temperature in the remote corona does not necessarily imply a high ablation efficiency. What matters for target acceleration is the amount of thermal energy that can be transported to the dense ablation region and converted into ablation pressure. The deeper deposition of the $3\omega$ component provides a more favorable thermal structure for this conversion.

\begin{figure}[htbp]
\centering
\includegraphics[width=1.00\textwidth]{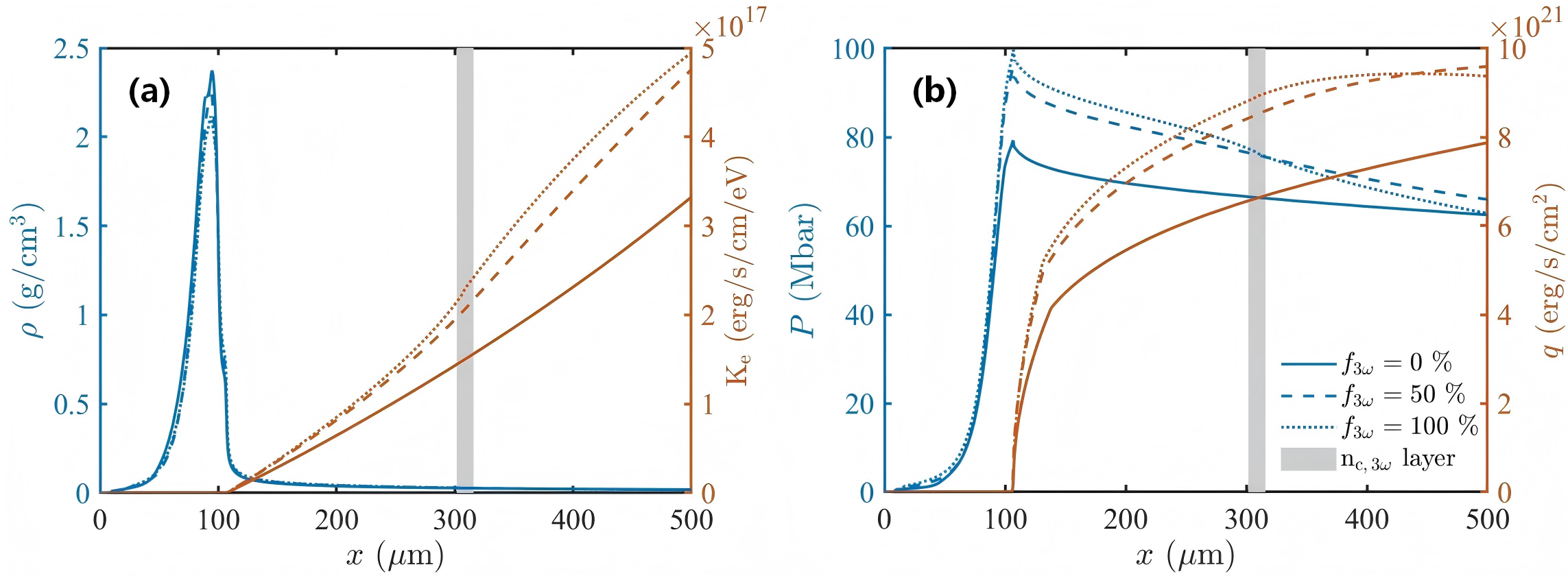}
\caption{
Thermal-transport profiles near the $3\omega$ critical layer. (a) Density and electron thermal conductivity $K_e$. (b) Ablation pressure $P_a$ and conductive heat flux $q$. The shaded region denotes the $3\omega$ critical layer. The mixed and pure $3\omega$ drives produce a larger thermal conductivity and heat flux near this region than the pure $2\omega$ drive, indicating more efficient conductive transport of thermal energy toward the dense ablation region.
}
\label{fig:transport_profile}
\end{figure}

\subsection{Thermal conductivity and heat flux}

To further clarify the energy-transport mechanism, Fig.~\ref{fig:transport_profile} compares the electron thermal conductivity, heat flux, and pressure profiles near the $3\omega$ critical layer. The shaded region marks the location of the $3\omega$ critical layer. The density profiles are also shown for reference. Figure~\ref{fig:transport_profile}(a) shows that the electron thermal conductivity $K_e$ near the $3\omega$ critical layer is significantly larger for the mixed and pure $3\omega$ drives than for the pure $2\omega$ drive. This behavior is consistent with the electron temperature structure in Fig.~\ref{fig:Te_profile}. Since the classical electron thermal conductivity depends nonlinearly on the electron temperature, approximately as
\begin{equation}
K_e \propto T_e^{5/2}
\end{equation}
in the Spitzer--H\"arm limit, even a moderate increase in the local electron temperature can lead to a substantial increase in the thermal conductivity. In the present simulations, the flux-limited conduction model modifies the classical transport in regions with steep temperature gradients. Nevertheless, the enhanced local electron temperature near the $3\omega$ critical layer still produces a larger effective thermal conductivity.

The enhanced conductivity leads to a stronger conductive heat flux. As shown in Fig.~\ref{fig:transport_profile}(b), the heat flux $q$ is larger for the mixed and pure $3\omega$ drives than for the pure $2\omega$ drive in the region near and beyond the $3\omega$ critical layer. This indicates that the inclusion of a $3\omega$ component opens a more effective thermal-conduction channel between the coronal plasma and the dense ablation region. As a result, the thermal energy stored in the corona can be transported inward more efficiently. The pressure profiles in Fig.~\ref{fig:transport_profile}(b) further support this interpretation. The mixed and pure $3\omega$ drives produce a higher pressure level near the dense plasma than the pure $2\omega$ drive.

\section{Discussion}
\label{sec:discussion}

The preceding sections show that the effect of the $3\omega$ component is not limited to a simple increase of the ablation pressure. Increasing $f_{3\omega}$ modifies the spatial distribution of laser-energy deposition, enhances the conductive heat flux $q$ toward the dense ablation region, increases the acceleration $g$, and raises the ablation velocity $V_a$. These changes reduce the target-incident laser energy required to reach a prescribed velocity and simultaneously lower the maximum linear ablative RTI gain. Therefore, the mixed $2\omega$--$3\omega$ drive provides a hydrodynamic design knob that couples drive efficiency and ablative stabilization. In this section, we combine the pressure-scaling, RTI-gain, and laser-energy results discussed above to identify a practical design window for the mixed-wavelength drive. Special attention is given to the equal-intensity mixed drive,
$$f_{3\omega}=50\%,$$
because this case captures much of the hydrodynamic benefit of pure $3\omega$ irradiation while retaining a substantial $2\omega$ component.

\subsection{Hydrodynamic trade-off between pure and mixed drives}

\begin{figure}[htbp]
    \centering
    \includegraphics[width=0.60\linewidth]{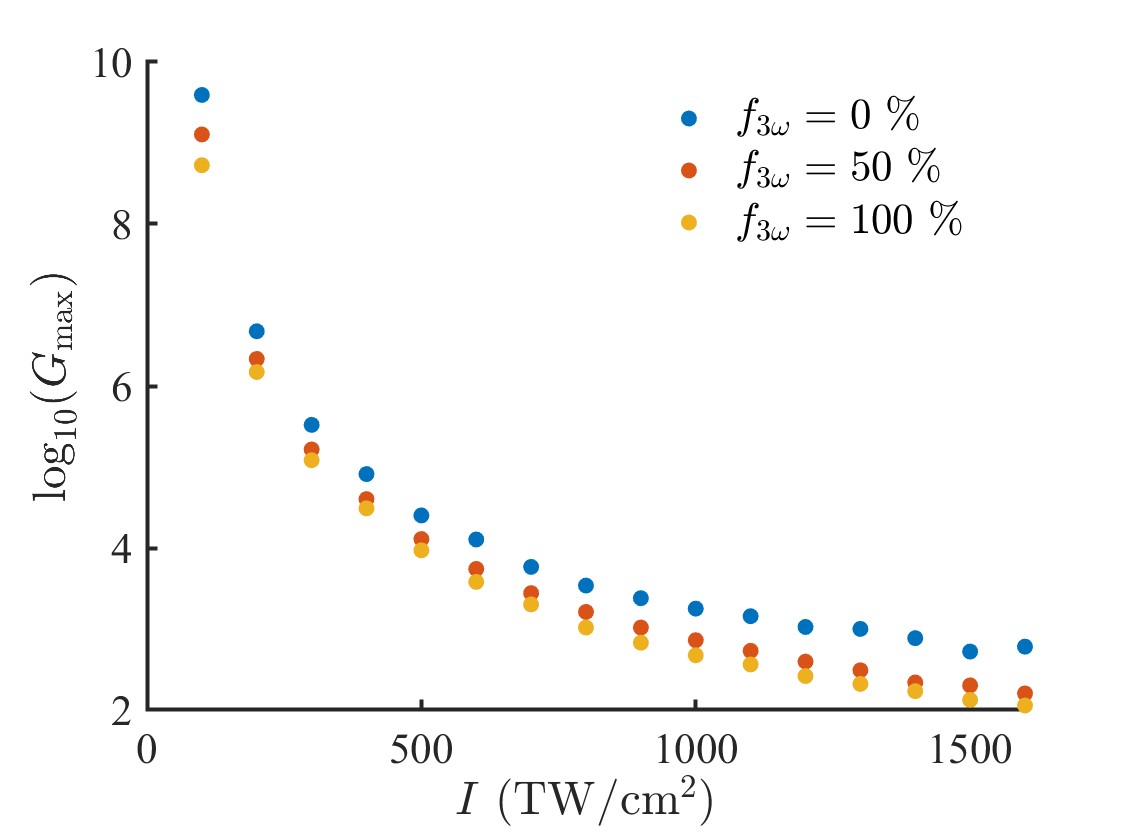}
    \caption{
    Maximum linear RTI gain for three representative drive configurations: pure $2\omega$ drive $(f_{3\omega}=0\%)$, equal-intensity mixed $2\omega$--$3\omega$ drive $(f_{3\omega}=50\%)$, and pure $3\omega$ drive $(f_{3\omega}=100\%)$. The equal-intensity mixed drive gives a substantially lower gain than pure $2\omega$ drive and remains close to pure $3\omega$ drive over the full intensity range.
    }
    \label{fig:design_gain}
\end{figure}

From the hydrodynamic point of view, the two limiting cases represent different levels of performance. Pure $2\omega$ drive requires the largest target-incident laser energy and gives the highest RTI gain in the present simulations. Pure $3\omega$ drive gives the lowest energy requirement and the lowest linear RTI gain. The equal-intensity mixed drive, $f_{3\omega}=50\%$, lies between these two limits, but its performance is much closer to pure $3\omega$ than to pure $2\omega$. This behavior is already reflected in the laser-energy comparison shown in Fig.~\ref{fig:energy_requirement}. At high intensity, such as $\sim 1600~\mathrm{TW/cm^2}$, the pure $2\omega$ drive requires substantially more target-incident laser energy than the pure $3\omega$ drive to reach $v_f=300~\mathrm{km/s}$. In contrast, the $f_{3\omega}=50\%$ case remains close to the pure $3\omega$ result over the intensity range considered here. Thus, adding only a partial $3\omega$ component is sufficient to recover most of the hydrodynamic efficiency of pure $3\omega$ irradiation. A similar conclusion can be drawn from the RTI-gain comparison. To highlight the design implication of the continuous $f_{3\omega}$ scan shown in Fig.~\ref{fig:rti_gain}(a), Fig.~\ref{fig:design_gain} replots three representative cases: pure $2\omega$ drive, equal-intensity mixed $2\omega$--$3\omega$ drive, and pure $3\omega$ drive. The plotted quantity is the maximum linear gain $G_{\max}$ obtained from the Takabe-type post-processing model. As shown in Fig.~\ref{fig:design_gain}, the $f_{3\omega}=50\%$ case substantially reduces $\log_{10}G_{\max}$ relative to pure $2\omega$ drive over the full intensity range. Although its gain is slightly higher than that of pure $3\omega$ at the same incident intensity, the difference between the $50\%$ mixed drive and pure $3\omega$ drive is much smaller than the difference between pure $2\omega$ and pure $3\omega$. This indicates that a partial $3\omega$ component captures a large fraction of the RTI-stabilizing effect of pure $3\omega$ irradiation.

\subsection{Compensation between mixing ratio and total laser intensity}

Another important implication of Fig.~\ref{fig:design_gain} is that the RTI penalty of using a partial rather than a full $3\omega$ drive can be compensated by a moderate increase in the total laser intensity. Since $G_{\max}$ decreases rapidly with $I$, increasing the intensity of the $f_{3\omega}=50\%$ drive by approximately $100$--$200~\mathrm{TW/cm^2}$ can bring its RTI gain close to that of the pure $3\omega$ drive at the original intensity. In some intensity ranges, the higher-intensity $f_{3\omega}=50\%$ case can even give a lower estimated RTI gain than the lower-intensity pure $3\omega$ case. This compensation can be expressed by an equivalent intensity offset $\Delta I_{\rm eq}$, defined implicitly by
\begin{equation}
    G_{\max}\left(I+\Delta I_{\rm eq}, f_{3\omega}=50\%\right)
    =
    G_{\max}\left(I, f_{3\omega}=100\%\right).
\end{equation}
For the parameter range considered here, $\Delta I_{\rm eq}$ is typically of the order of
\begin{equation}
    \Delta I_{\rm eq} \simeq 100\text{--}200~\mathrm{TW/cm^2}.
\end{equation}
Thus, the total laser intensity and the mixing ratio can be traded against each other in designing a mixed-wavelength drive. A moderate increase in target-incident intensity may allow an equal-intensity mixed drive to reach a linear RTI comparable to that of pure $3\omega$, while requiring only half of the laser intensity to be carried by the $3\omega$ component.

\subsection{Design implication of the $f_{3\omega}=50\%$ mixture}

The equal-intensity mixed drive is therefore a particularly attractive point in the present hydrodynamic design space. In terms of laser-energy requirement, $f_{3\omega}=50\%$ remains close to the pure $3\omega$ limit and recovers most of the hydrodynamic efficiency lost in pure $2\omega$ irradiation. In terms of RTI, it gives a much lower gain than pure $2\omega$, while being only moderately above the pure $3\omega$ result at the same intensity. If a small increase in the total target-incident intensity is allowed, the $f_{3\omega}=50\%$ case can reach a linear RTI comparable to that of pure $3\omega$. The physical origin of this favorable behavior is the two-region energy-deposition structure of the mixed drive.

However, several limitations should be emphasized. First, the present energy comparison is based on the target-incident laser energy in the radiation-hydrodynamic simulations. It does not include facility-level frequency-conversion efficiency, beam-transport loss, or the energy cost of generating the $2\omega$ and $3\omega$ components. Therefore, the energy reduction discussed here should be interpreted as an improvement in target hydrodynamic efficiency rather than a complete facility-level energy balance. Second, LPIs are not included. This is particularly important for mixed-wavelength drive because the $2\omega$ and $3\omega$ components may have different susceptibility to stimulated Brillouin scattering, stimulated Raman scattering, two-plasmon decay, etc. These processes can modify absorption, preheat, coronal plasma conditions, and ultimately the RTI-relevant hydrodynamic parameters. The present conclusions therefore represent a hydrodynamic assessment rather than a full evaluation of experimental viability. Third, the RTI analysis is based on 1D hydrodynamic profiles and a linear Takabe-type post-processing model. Multidimensional effects such as laser imprint, feedthrough, nonlinear bubble-spike evolution, and spherical convergence are not included. The values of $G_{\max}$ should therefore be interpreted as comparative linear stability metrics rather than absolute predictions of perturbation amplitudes in an implosion experiment. Despite these limitations, the present results identify a clear hydrodynamic mechanism and a useful design principle: introducing a moderate $3\omega$ fraction can improve the spatial coupling of laser energy to the ablation region, increase $V_a$, reduce the required target-incident energy, and lower the estimated linear ablative RTI gain. Future work should combine this hydrodynamic optimization with multidimensional radiation-hydrodynamic simulations and laser--plasma-interaction modeling.

\section{Conclusion}
\label{sec:conclusion}

In this work, we have performed 1D radiation-hydrodynamic simulations of planar CH targets driven by mixed $2\omega$--$3\omega$ lasers. The total target-incident intensity was varied from $100$ to $1600~\mathrm{TW/cm^2}$, and the $3\omega$ intensity fraction was scanned from pure $2\omega$ to pure $3\omega$. Thick CH targets were used to characterize quasi-steady ablation, while thin CH foils were used to evaluate acceleration-stage hydrodynamics, linear ablative RTI gain, and the target-incident energy required to reach $300~\mathrm{km/s}$. The ablation pressure follows a power-law dependence on the incident intensity, $P_a = A I^m$. In the thick-target quasi-steady regime, the exponent remains nearly insensitive to the wavelength mixture, with $m\simeq 0.80$--$0.82$, while the normalization $A$ increases monotonically with $f_{3\omega}$. In the finite-foil acceleration stage, rarefaction and finite-target effects reduce both the pressure normalization and the intensity exponent, giving $m_{\mathrm{acc}}\simeq 0.68$--$0.77$. The pressure enhancement associated with $3\omega$ mixing is considerably stronger during foil acceleration than in the thick-target ablation regime, reaching about $34\%$ for pure $3\omega$ relative to pure $2\omega$ at $I=1600~\mathrm{TW/cm^2}$.

A Takabe-type post-processing model was used to estimate the maximum linear ablative RTI gain under a fixed final-velocity constraint. The analysis shows that, in the limit $L_m=0$, increasing the acceleration alone does not reduce the gain envelope because the larger instantaneous growth rate is compensated by the shorter acceleration time. For finite $L_m$, the gain envelope decreases with increasing acceleration, while the most direct stabilizing effect comes from the increase of the ablation velocity $V_a$. The finite density-gradient scale length further suppresses high-wavenumber modes. When the hydrodynamic quantities extracted from the FLASH simulations are used in this model, the estimated maximum linear RTI gain decreases with both increasing laser intensity and increasing $f_{3\omega}$. This reduction is primarily caused by the enhanced $V_a$, the increased $g$, with additional contributions from finite-$L_m$ stabilization.

Adding a $3\omega$ component also lowers the incident laser energy required to accelerate the foil to $300~\mathrm{km/s}$. Pure $2\omega$ drive requires the largest target-incident energy, whereas pure $3\omega$ drive gives the lowest value. The equal-intensity mixed case, $f_{3\omega}=50\%$, remains close to the pure $3\omega$ result over most of the intensity range. At high intensity$(\sim 1600~\mathrm{TW/cm^2})$, pure $2\omega$ requires about $40\%$ more target-incident energy than pure $3\omega$, while the $50\%$ mixed case is only a few percent $(3\sim5\%)$ above the pure $3\omega$ value. The improvement is traced to the spatial redistribution of laser-energy deposition. Because $n_c \propto \omega^2$, the $3\omega$ component deposits energy in a denser region closer to the ablation front than the $2\omega$ component. This produces a more favorable electron-temperature and thermal-conductivity profile near the dense plasma, enhances the conductive heat flux toward the ablation region, and increases the acceleration-stage ablation pressure. The same mechanism raises the acceleration, the ablation velocity, and thereby strengthens ablative RTI stabilization.

These results indicate that mixed $2\omega$--$3\omega$ irradiation provides an additional hydrodynamic design degree of freedom for direct-drive targets. In the present 1D planar model, the equal-intensity mixture is a representative compromise: it recovers much of the hydrodynamic efficiency and linear RTI stabilization of pure $3\omega$ drive while retaining a substantial $2\omega$ component. However, the present study does not include facility-level frequency-conversion efficiency, multidimensional perturbation growth, spherical convergence, laser--plasma instabilities, etc. Future work should combine mixed-wavelength hydrodynamic optimization with multidimensional radiation-hydrodynamic simulations and LPIs modeling to determine the full design window for experimentally relevant direct-drive implosions.

\section*{Acknowledgments}
\label{Acknowledgments}

This work was supported by the Science Challenge Project (No. TZ202514), National Natural Science Foundation of China (No. 12375242).

\bibliography{Manuscript.bib}
	
\end{document}